\documentclass{article}

\usepackage{arxiv}

\usepackage[utf8]{inputenc} 
\usepackage[T1]{fontenc}    
\usepackage{hyperref}       
\usepackage{url}            
\usepackage{booktabs}       
\usepackage{amsfonts}       
\usepackage{nicefrac}       
\usepackage{microtype}      
\usepackage{lipsum}		
\usepackage{xcolor} 
\usepackage{graphicx}
\usepackage{natbib}
\usepackage{doi}
\usepackage{amsmath}
\newcommand{\dd}{\mathrm{d}}

\title{Evaluating consumption effects of intelligent control algorithms for district heated buildings}

\author{{Antti Solonen}\\
	Danfoss Leanheat \\ 
        LUT University\\
	\texttt{antti.solonen@danfoss.com} \\
	\AND
	Arttu Häkkinen \\
        Danfoss Leanheat \\
        LUT University \\
	\And
	Sallamaari Rapo \\
	Danfoss Leanheat \\
    \And
	Antti Mäkinen \\
	Danfoss Leanheat \\
    \And
	Sampo Kaukonen \\
	Danfoss Leanheat \\
    \And
	Felipe Uribe \\
	Danfoss Leanheat \\
}

\renewcommand{\shorttitle}{Evaluating intelligent control effects}

\hypersetup{
pdftitle={Evaluating intelligent control effects},
pdfsubject={},
pdfauthor={Antti Solonen},
pdfkeywords={First keyword, Second keyword, More},
}

\begin{document}
\maketitle

\begin{abstract}
As buildings become increasingly connected and sensor-rich, intelligent remote heating control is rapidly superseding conventional local heating control. Such control algorithms often aim at reducing energy consumption by minimizing over-heating and utilizing free solar energy, for instance. Numerous companies offering heating optimization solutions have recently emerged. After installing such a system, end-users naturally want to quantify and verify the effect of such an investment, i.e., monetary return. Methods for tracking buildings' heating efficiency are diverse, ranging from simple weather normalization to more complex modeling approaches, but lack transparency and commonly agreed best practices. The problem is further complicated by the fact that buildings constantly undergo non-control-related changes that affect their energy efficiency, making it difficult to isolate and track only control-related effects using the existing methods.

In this paper, we first review and derive methods for monitoring the overall efficiency of buildings, and show their inability to isolate the control effects from other changes happening in the buildings. We then propose a model-based approach for estimating and tracking \textit{only} the control-related effects. Moreover, we show how the models can \textit{decompose} the total control effect into sub-components to reveal where the energy effects come from. We demonstrate the methods using real data collected over approximately 10 years from the Danfoss Leanheat Building platform. Our scope focuses on district heated buildings with substation-level (supply temperature) control, but the methodology extends to other cases as well.
\end{abstract}


\section{Introduction}
The building sector is responsible for approximately 32\% of global energy use and 34\% of global $\text{CO}_2$ emissions \citep{UNEP2025}. To achieve climate goals, significant steps must be taken to reduce the rapidly growing energy consumption in this sector \citep{iea_2025}, affecting both new construction and the transformation of the current building stock. In response to this challenge, intelligent control solutions for district heated buildings have become popular in recent years due to the increased availability of data and IoT solutions for remote controllability \citep{yao_and_shekhar_2021, taheri_et_al_2022}. These solutions typically replace the traditional \textit{open loop} heating curve control with \textit{closed loop} control that receives on-line feedback from installed indoor temperature sensors. These methods typically aim at reducing energy consumption in the buildings by avoiding the over-heating that often occurs with heating curve control, especially when free energy from solar radiation is available, for instance. See, e.g., \cite{drgona2020all, hannula26, hsu_2025, ma_2025} for recent discussions on model predictive control approaches for heating.

As an increasing number of companies providing heating control optimization services have emerged, a natural question arises from end users: what was the effect of installing an intelligent control system in my building and how profitable was my investment? This calls for methodology to assess consumption changes in buildings. Various methods are available for such calculations, ranging from weather normalization methods to supply temperature based approaches and advanced modeling techniques. However, as discussed in Section \ref{sec:literature}, most existing methods aim at quantifying and tracking changes in the \textit{total heat consumption} of the building, and fail to isolate the specific contribution of the intelligent control system. There are changes constantly happening in buildings that affect consumption, such as major renovations, changes to the ventilation system and slow performance degradation due to aging of the building and equipment. To get a realistic picture of the contribution of intelligent control alone, these non-control-related effects must be somehow eliminated from the analysis. This problem is especially pronounced for buildings that have been in intelligent control for a long time and have thus undergone slow changes and sudden shifts that have likely affected their overall consumption levels independent of the heating control method.

The goal of this paper is two-fold. First, we discuss various existing methods for overall evaluation of building heating efficiency to bring clarity and transparency to the benefits and downsides of the different approaches. Second, we propose a novel modeling-based technique for separating control-related effects from all other consumption changes occurring in the buildings. We further show how such an approach can be used to decompose the control effect into various sub-components such as the contributions of solar radiation and chosen indoor temperature levels. Our scope is district heated buildings with substation-level control, which assumes that indoor temperatures are directly controllable by modifying supply temperature values at the substation, though the approach extends to other control modes, such as thermostat-based room-level control. We evaluate the methodologies using real anonymized data collected from multi-family buildings over the last decade via the Danfoss Leanheat Building (LHB) platform\footnote{\url{https://www.danfoss.com/en/products/dhs/software-solutions/danfoss-leanheat-software-suite-services/leanheat-building/}}.

The paper is organized as follows. Section \ref{sec:literature} contains a literature review of the existing efficiency evaluation methods, and Section \ref{sec:methods} presents various methods in more mathematical detail. Section \ref{sec:changes} discusses various non-control-related consumption changes and how we can detect them from available data. Section \ref{sec:lheffect} then shows how we can use models to target only the control-related changes and how we can decompose the effect into sub-components. Section \ref{sec:conclusions} concludes the paper.


\section{Literature review and novelty}
\label{sec:literature}


Most of the existing literature for heating performance evaluation of buildings focuses on weather normalization methods that aim to remove the impact of weather on building consumption data, thus making data from different time periods comparable. As discussed in Section \ref{sec:normalization}, weather normalization methods are typically based on the concept of Heating Degree Days (HDDs), which appear in the literature as early as in \cite{thom54}. A detailed literature review of HDD-based normalization methods is given in \cite{rapo25}. Weather normalization methods are used to evaluate the total consumption and thus cannot isolate control-related effects, which is one of the main topics of this paper.

Globally recognized Measurement and Verification (M\&V) frameworks, such as \cite{evo2012ipmvp} and \cite{ashrae2014guideline}, provide standardized options for performance tracking. Furthermore, the accuracy of the involved M\&V baseline models has been analyzed in, e.g., \cite{mathieu2011quantifying, granderson2016accuracy,alrobaie_and_krarti_2023}. However, since these weather normalization methods aim to evaluate the total heat consumption of buildings, they are not usable for isolating the control-related effects.

In a recent work, \cite{saloux25} discusses and compares various model-free and model-based performance evaluation techniques in the case of a single building and finds model-based methods more suitable for performance tracking. Similarly as in weather normalization methods, the discussed model-based techniques focus on evaluating the changes in total heat consumption, whereas the main interest of this work is to isolate the control-related effects from the other performance changes. To the best of our knowledge, this problem has not been studied before, although \cite{saloux25} acknowledges the problem that savings estimates may incorporate effects not related to the control strategy. 

In principle, our method becomes close to the two-model strategy (approach 4) proposed in \cite{saloux25}, where models are developed for both before and after an event (intelligent control activation), which allows extrapolating savings estimates to future times. We build upon this work by developing gray-box models that include all relevant control-related parameters and also allow decomposing the savings estimates into sub-components.

Overall, tracking the total performance of buildings is a relatively well-studied field. However, the literature lacks a comprehensive discussion and derivation of the different methods and, as shown in the following sections, the existing methods fail to isolate control related changes from other consumption changes happening in the buildings. This paper aims at bridging the gap by introducing the following key contributions:
\begin{itemize}
    \item[(i)] \textbf{Reviewing and deriving the existing performance tracking methods.} We derive a number of overall performance tracking methods and discuss the assumptions, benefits and downsides of each. Moreover, building consumption models are discussed, as they provide the backbone of various performance tracking approaches. Supply temperature -based methods are introduced as a novel option in addition to the existing model-based methods and weather normalization techniques.
    \item[(ii)] \textbf{Demonstrating and diagnosing non-control-related performance changes from data.} We show what kind of consumption changes can be seen in the buildings and discuss how these changes can be diagnosed based on collected data. Such diagnostics can be applied to develop useful consumption analytics tools. We demonstrate the consumption changes using consumption data from real buildings.
    \item[(iii)] \textbf{Presenting a method for isolating the control-related effects.} We show that the existing methods track the total performance but fail to capture only the control-related consumption changes. We propose a simulation-based method for separating these control-related changes from the other changes. The algorithm is demonstrated using consumption data from real buildings.
    \item[(iv)] \textbf{Presenting a method for decomposing the control-related consumption change into sub-components.} We show how the proposed simulation-based technique can be used to divide the control-related changes into sub-components such as solar radiation gains and the effect of chosen indoor temperature levels. This decomposition yields interesting insights into the sources of energy savings obtained through intelligent control.
\end{itemize}


\section{Overall efficiency evaluation methods}
\label{sec:methods}

\subsection{Modeling heating power}


All performance tracking methods are essentially based on mathematical models of the heating power. Models are needed to remove the effects of weather conditions and other variables, thereby enabling meaningful comparison of power readings at different times with varying conditions. Here, we present a simple derivation of a heating power model, based largely on \cite{rapo25} and \cite{hakkinen23}, and then discuss in the following sections how the models can be used in different efficiency evaluation methods. 

We discuss two different types of models. First, we present a model that maps weather conditions to power consumption, which is usable in model-based efficiency tracking and weather normalization approaches described in sections \ref{sec:modelbased} and \ref{sec:normalization}, respectively. Then, we discuss a model that maps supply temperature to power consumption, which is needed in the supply temperature based performance tracking methods, which are further discussed in Section \ref{sec:supplytemp}.

\subsubsection{Weather vs. heating power}

\begin{figure}[!ht]
    \centering
    \includegraphics[width=0.5\linewidth]{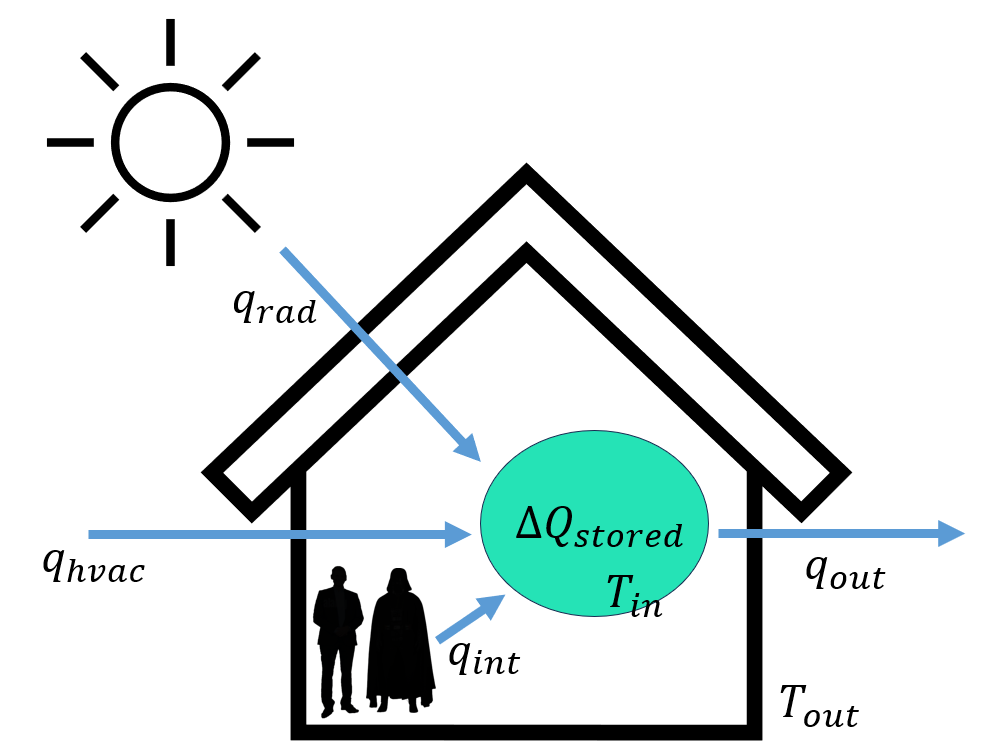}
    \caption{Energy flows in and out from a building: $q_{\mathrm{hvac}}$ denotes the heat flowing in from the heating system, $q_{\mathrm{rad}}$ is the contribution of the solar radiation, $q_{\mathrm{int}}$ are the internal heat gains from, e.g., people and appliances, and $q_{\mathrm{out}}$} is the heat flow out of the building through the building envelope. Indoor and outdoor temperatures are denoted by $T_{\mathrm{in}}$ and $T_{\mathrm{out}}$, respectively. The figure is reproduced with author's permission from \cite{rapo25}.
    \label{fig:heatflows}
\end{figure}

A model of the heating power of a building can be derived using the law of energy conservation. The rate of change of energy stored in the building, denoted here by $Q_{\mathrm{stored}}$, must equal the sum of incoming energy flows (heat provided by the heating system, solar radiation and internal heat sources) minus the heat losses to the surroundings. See Fig.~\ref{fig:heatflows} for an illustration and the notation of the different energy flows. 

The energy conservation law can then be written as
\begin{equation}
    \label{eq:energy_balance}
    \frac{\dd Q_{\mathrm{stored}}}{\dd t} = q_{\mathrm{hvac}} + q_{\mathrm{rad}} + q_{\mathrm{int}} - q_{\mathrm{out}}.
\end{equation}

Solving Eq.~\eqref{eq:energy_balance} for the required heating power yields
\begin{equation}
    \label{eq:hvac_power}
    q_{\mathrm{hvac}} = \frac{\dd Q_{\mathrm{stored}}}{\dd t} + q_{\mathrm{out}} - q_{\mathrm{rad}} - q_{\mathrm{int}},
\end{equation}
that is, part of the power goes to changing the energy stored in the system, heat losses to the surroundings $q_{\mathrm{out}}$ increase the required heating power and the gains from solar radiation and internal sources, $q_{\mathrm{rad}}$ and $q_{\mathrm{int}}$, reduce the required power.

Eq.~\eqref{eq:hvac_power} is not yet applicable for practical use, we need to express the energy flows with more useful quantities first. Expressing the stored energy in the indoor air of the building yields $Q_{\mathrm{stored}} = C_{\mathrm{in}}T_{\mathrm{in}}$, where $C_{\mathrm{in}}$ is the heat capacity of the building. The heat loss to surroundings can be assumed to be proportional to the temperature difference between indoor and outdoor air. The solar radiation effect is difficult to model accurately; the simplest useful model states that the heating power from solar radiation is proportional to the total solar radiation intensity, denoted here by $\Phi_{\mathrm{rad}}$. With these choices, a candidate model for heating power can be written as
\begin{equation}
\label{eq:simple}
    q_{\mathrm{hvac}} = C_{\mathrm{in}} \frac{\dd T_{\mathrm{in}}}{\dd t} + b\left( T_{\mathrm{in}} - T_{\mathrm{out}}\right) - c\Phi_{\mathrm{rad}} - q_{\mathrm{int}},
\end{equation}
where $b$ and $c$ are unknown constants. This formulation effectively acts as a simplified thermal resistance-capacitance (RC) network or a gray-box model. The suitability and parameter identification challenges of such models for building heat dynamics have been well established in the literature, see, e.g., \cite{bacher2011identifying, reynders2014quality}.

The internal heat gains $q_{\mathrm{int}}$ are hard to model from first principles. Our goal is to model the power at daily resolution (and even lower with some weather normalization approaches), which is enough for performance tracking purposes. While $q_{\mathrm{int}}$ typically shows clear temporal patterns within the hours of a day, at daily resolution we can assume that the term is roughly constant and thus treat $q_{\mathrm{int}}$ as another unknown coefficient in the model.

Note that in general we do not get data for $q_{\mathrm{hvac}}$ alone, but instead we measure the total heat consumption $q_{\mathrm{tot}}$ that also includes domestic hot water (DHW) consumption $q_{\mathrm{dhw}}$. Adding this to the model in Eq.~\eqref{eq:simple} yields
\begin{equation}
\label{eq:simpletot}
    q_{\mathrm{tot}} = q_{\mathrm{hvac}} + q_{\mathrm{dhw}} = C_{\mathrm{in}} \frac{\dd T_{\mathrm{in}}}{\dd t} + b\left( T_{\mathrm{in}} - T_{\mathrm{out}}\right) - c\Phi_{\mathrm{rad}} - q_{\mathrm{int}} + q_{\mathrm{dhw}}.
\end{equation}

As in $q_{\mathrm{int}}$, DHW consumption depends on human behavior and is thus hard to model physically. It also has clear temporal patterns within the hours of a day (see, e.g., \cite{jordan2001influence}), but in daily resolution we can treat it as a constant. Note, however, that $q_{\mathrm{int}}$ and $q_{\mathrm{dhw}}$ cannot be separately estimated when the model is calibrated, so we have to settle for expressing them as a combined unknown constant $q_0=q_{\mathrm{dhw}}-q_{\mathrm{int}}$.

With these assumptions, Eq.~\eqref{eq:simpletot} provides the first model where the output and all inputs are available from different data sources: $q_{\mathrm{tot}}$ from energy meters, $T_{\mathrm{in}}$ from indoor temperature sensors (if available), and the weather inputs $T_{\mathrm{out}}$ and $\Phi_{\mathrm{rad}}$ from weather data providers and/or local outdoor sensors. Calibrating the unknown coefficients $\theta = \left[ C_{\mathrm{in}}, b, c, q_0 \right]$ using the available data amounts to standard linear regression.

\begin{figure}[!ht]
    \centering
    \includegraphics[width=0.8\linewidth]{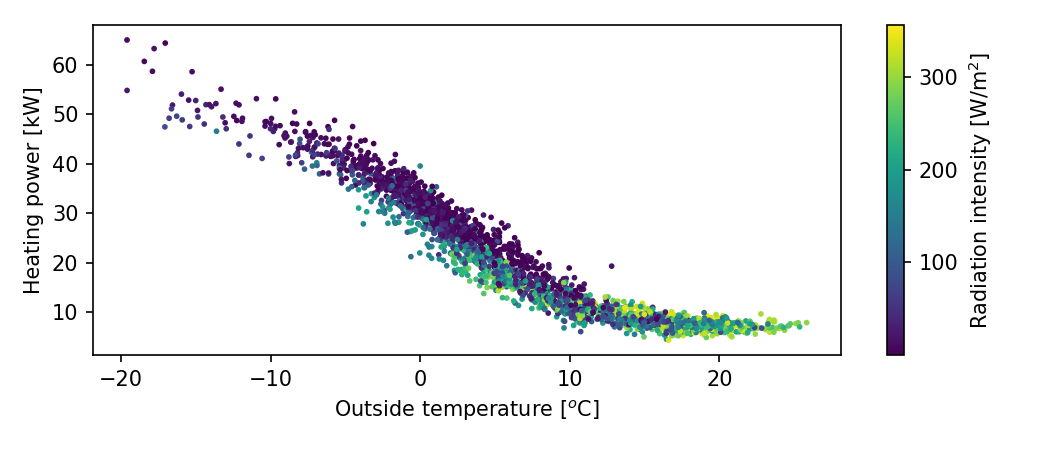}
    \caption{Daily average power consumption vs. outside temperature, color indicates the solar radiation intensity.}
    \label{fig:tout_vs_power}
\end{figure}

Although the simple model above is useful as such, its accuracy can be greatly improved in a few simple ways. To see this, let us take a look at Fig.~\ref{fig:tout_vs_power}, where the daily average power is plotted against outside temperature and solar radiation intensity for a selected building. Some observations can be made that are not properly captured by the above simple model. First, the cooling effect is outdoor temperature dependent: when it gets warm outside, the space heating demand eventually disappears, leaving only the DHW demand. Additionally, at cold temperatures the slope between $T_{\mathrm{out}}$ and $q_{\mathrm{hvac}}$ changes slightly. Second, the solar radiation effect is also outdoor temperature dependent: at warm outdoor temperatures, when the space heating demand diminishes, the solar radiation cannot anymore be fully utilized because the supply temperature is already lowered to its minimum value.

\begin{figure}[!ht]
    \centering
    \includegraphics[width=0.8\linewidth]{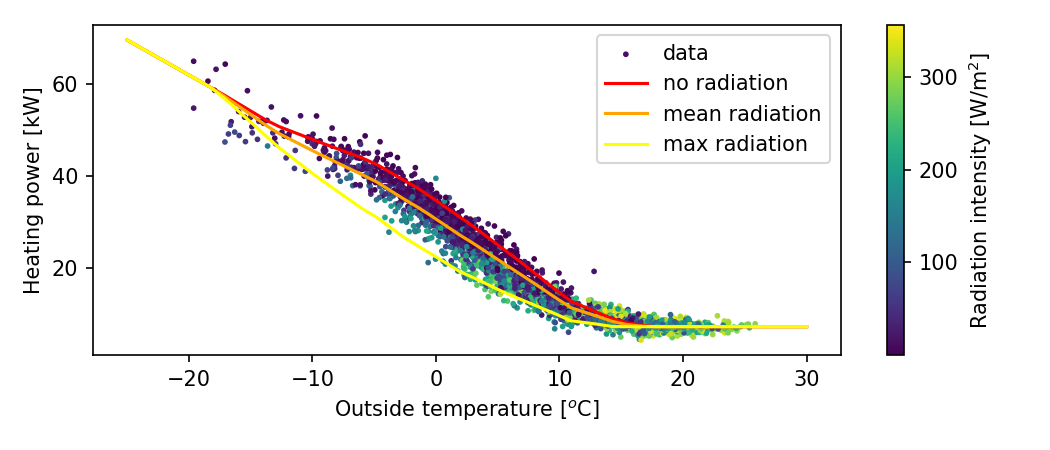}
    \caption{Daily average power consumption vs. outside temperature, color indicates the solar radiation intensity. Lines indicate model predictions with different solar radiation intensities. In the model predictions, indoor temperature was fixed to its mean value.}
    \label{fig:gam_fit}
\end{figure}

These observations motivate a more general heating power model that can capture the nonlinearities. Expressing the cooling term and the radiation effect coefficient as functions instead of constants yields
\begin{equation}
\label{eq:gam}
    q_{\mathrm{tot}} = C_{\mathrm{in}} \frac{\dd T_{\mathrm{in}}}{\dd t} + f_1\left( T_{\mathrm{in}} - T_{\mathrm{out}}\right) - f_2(T_{\mathrm{out}})\Phi_{\mathrm{rad}} + q_0,
\end{equation}
where $f_1$ and $f_2$ are now unknown functions. Instead of imposing a parametric expression for the functions, we can treat them in a non-parametric way by expressing them using chosen basis functions. We also know something about their behavior: (i) their values approach zero when the outdoor temperature becomes sufficiently high, and (ii) their behavior is likely smooth (e.g., the cooling effect does not change significantly when the outdoor temperature changes slightly). Encoding these \textit{prior} assumptions into the model leads naturally to Bayesian Generalized Additive Model (GAM) formulations, discussed, for example, in \cite{solonen23}. Data and model predictions with various solar radiation intensities for such a GAM are illustrated in Fig.~\ref{fig:gam_fit}: one can see that the discussed nonlinearities are well captured. For brevity, we omit the implementation details of such a GAM formulation here; some of these can be found in Appendix~\ref{sec:appendix_A}.

\subsubsection{Supply temperature vs. heating power}
\label{sec:te1_vs_pow}

\begin{figure}[!ht]
    \centering
    \includegraphics[width=0.9\linewidth]{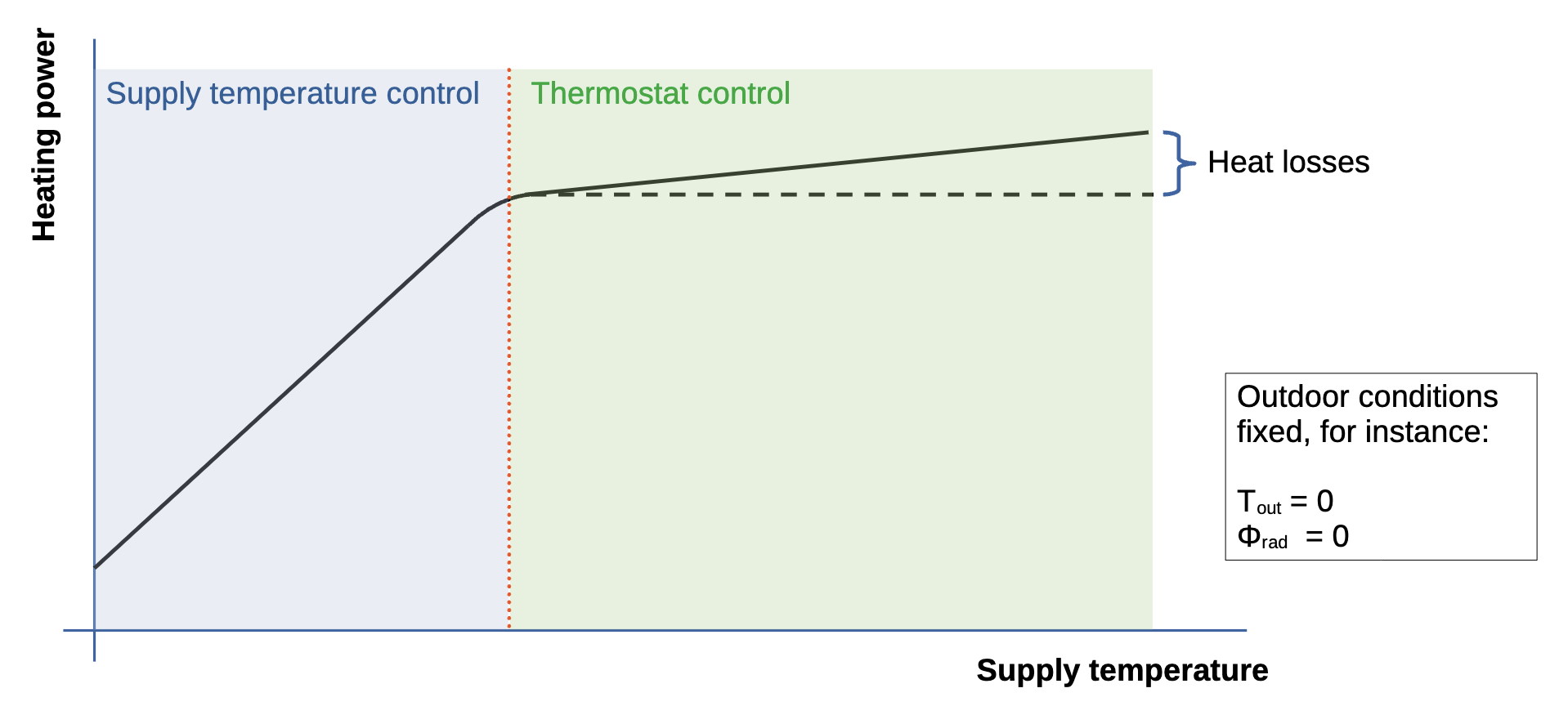}
    \caption{Illustration of two ``control regimes'' in fixed outdoor conditions. If supply temperature is lower than a (outdoor condition specific) critical value, we are in the ``supply temperature control regime'' where the supply temperature directly dictates the heating power and indoor comfort. In contrast, the right side of the critical value is the ``thermostat regime'' where thermostats regulate the flow and dependency on supply temperature is weak.}
    \label{fig:regimes}
\end{figure}

For supply temperature based control effect calculations, discussed in more detail in Section \ref{sec:supplytemp} below, we need a model that maps power consumption to the secondary side supply temperature. The relationship between supply temperature and heating power depends largely on which ``control regime'' we are at: one where indoor temperatures and heating power are directly affected by the supply temperatures (with thermostats not significantly controlling radiator flow), or another one called the ``thermostat regime'' where thermostats heavily regulate the consumption. These regimes are illustrated in Fig.~\ref{fig:regimes} for fixed outdoor conditions: below a certain supply temperature value, thermostats are inactive and the heating power reacts strongly to supply temperature. Above this critical value, thermostats are active and heating power depends weakly on supply temperature due to increased heat losses at higher supply temperatures.

In the ``thermostat regime'', the dependency between heating power and supply temperature is weak, and heating power is dictated by the chosen indoor temperature level in the thermostats, as described by Eq.~\eqref{eq:gam}. Here, we assume here that we are in the ``supply temperature control regime''. Deriving the power equation for mapping supply temperature to heating power in this regime starts from the energy balance: heating power in the space heating circuit must match the difference between power entering the system (via supply temperature) and power leaving the system (via return temperature):
\begin{equation}
\label{eq:deltaT_0}
    q_{\mathrm{tot}} = \dot{m} c_p \left( T_{\mathrm{sup}}-T_{\mathrm{ret}} \right) + q_{\mathrm{dhw}},
\end{equation}
where $\dot{m}$ is the mass flow rate through the radiators, $c_p$ is the heat capacity of water, $T_{\mathrm{sup}}$ is the supply temperature and $T_{\mathrm{ret}}$ is the return temperature.

For the model to be applicable in performance tracking applications, we need to simplify it by eliminating the return temperature $T_{\mathrm{ret}}$ from Eq.~\eqref{eq:deltaT_0}. Modeling the radiator as a simple pipe that emits heat into its surroundings yields
\begin{equation}
\label{eq:deltaT}
    T_{\mathrm{ret}} = T_{\mathrm{in}} + \mathrm{exp} \left( -\frac{\alpha}{\dot{m}} \right) \left( T_{\mathrm{sup}} - T_{\mathrm{in}} \right),
\end{equation}
where $\alpha$ is a building-specific unknown constant. We refer to \cite{hakkinen23} for the full derivation of this formula. Now, substituting Eq.~\eqref{eq:deltaT} into \eqref{eq:deltaT_0} and assuming that the flow rate through the radiators is constant, we can write
\begin{equation}
\label{eq:pow_vs_sup}
    q_{\mathrm{tot}} = k \left( T_{\mathrm{sup}}-T_{\mathrm{in}} \right) + q_{\mathrm{dhw}},
\end{equation}
where $k$ and $q_{\mathrm{dhw}}$ are building-specific coefficients that can be estimated from data. We now have a simple mapping from supply temperature to total heating power, which can be used in the supply temperature based control effect calculations, discussed in Section \ref{sec:supplytemp}.

\begin{figure}[!ht]
    \centering
    \includegraphics[width=0.8\linewidth]{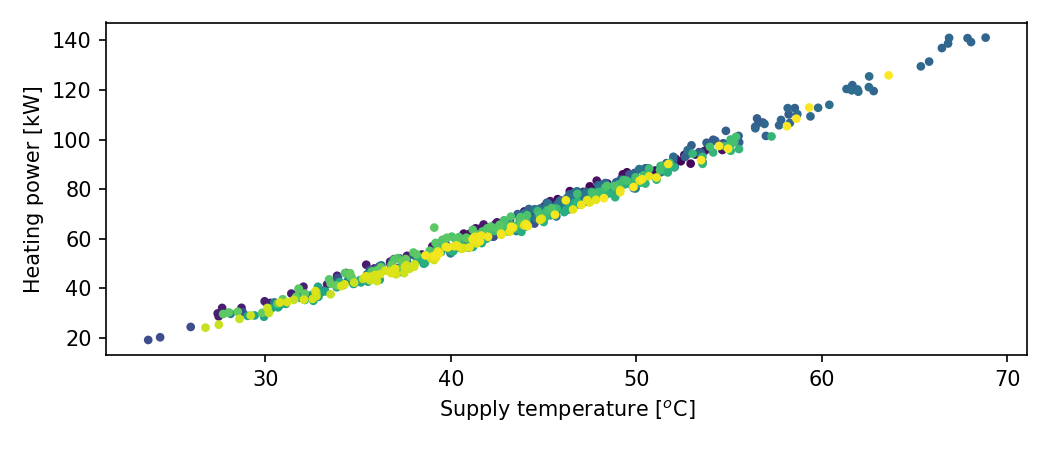}
    \caption{Total heating power vs. supply temperature (daily averages), color indicates time.}
    \label{fig:te1_vs_power}
\end{figure}

Note that theoretically this derivation applies only to a single radiator, but we have observed that such a model works well also at the building scale. In Fig.~\ref{fig:te1_vs_power} we show an example of the relationship between supply temperature and total heating power, demonstrating that the relationship is indeed linear. Moreover, we see no large deviations from this linear behavior, indicating that the system is likely in the supply temperature control regime. This is also our general observation in the Nordic countries: for most buildings, the relationship between supply temperature and heating power appears to be linear without significant ``thermostat effects''. 

\subsection{Model-based efficiency tracking}
\label{sec:modelbased}

Perhaps the most straightforward way to apply the above models for evaluating the effect of an event in a heating system, such as installing an intelligent control solution, is to perform a \textit{before-after} comparison. We can, for instance, characterize the consumption behavior of the system under different weather conditions by calibrating a power model using data from before the event. Then, we can simulate \textit{what the power consumption would have been} with the old setup and compare the model simulations to realized power readings after the event. 

More formally, let us denote the ``pre-event'' baseline model as $q_{\mathrm{baseline}} = f(T_{\mathrm{out}}, \Phi_{\mathrm{rad}})$. The model could be, for instance, some version of Eq.~\eqref{eq:gam}\footnote{Note, however, that indoor temperature $T_{\mathrm{in}}$ is not often available before installing an intelligent control solution, but assuming constant $T_{\mathrm{in}}$ often yields accurate enough models, since changes in $T_{\mathrm{in}}$ are small compared to the $T_{\mathrm{out}}$ effect.}. Then, we can assess efficiency changes at the building by comparing the simulated pre-event consumption during the observed period to the observed total power consumption $q_{\mathrm{tot}}^{\mathrm{obs}}$:
\begin{equation}
    \label{eq:modelbased}
    \mathrm{effect\,[\%]} = 100 \cdot \frac{q_{\mathrm{tot}}^{\mathrm{obs}}-f(T_{\mathrm{out}}^{\mathrm{obs}}, \Phi_{\mathrm{rad}}^{\mathrm{obs}})}{f(T_{\mathrm{out}}^{\mathrm{obs}}, \Phi_{\mathrm{rad}}^{\mathrm{obs}})},
\end{equation}
where $T_{\mathrm{out}}^{\mathrm{obs}}$ and $\Phi_{\mathrm{rad}}^{\mathrm{obs}}$ are the observed outdoor temperature and solar radiation intensity, respectively. 

\begin{figure}[!ht]
    \centering
    \includegraphics[width=1.0\linewidth]{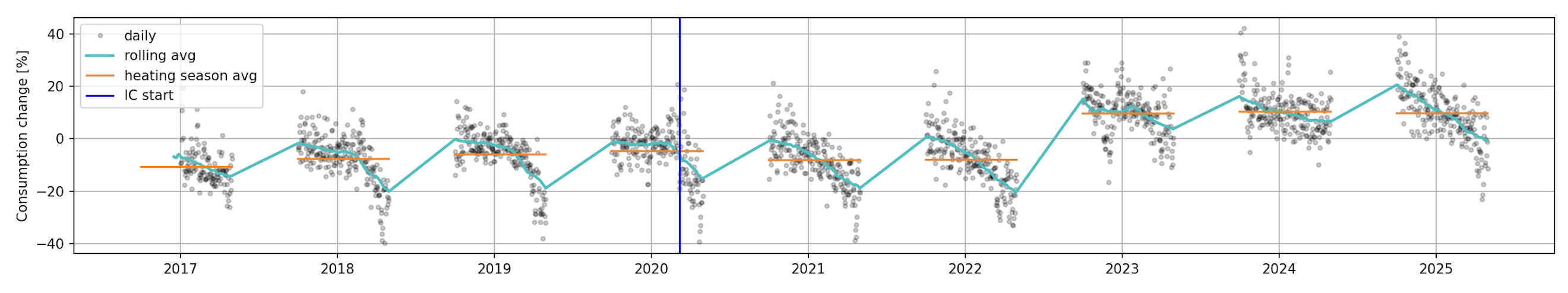}
    \caption{Model-based efficiency tracking for an example building.}
    \label{fig:perf_tracking}
\end{figure}

Eq.\eqref{eq:modelbased} yields a normalized efficiency time series for the building. An example of such an analysis is given in Fig.~\ref{fig:perf_tracking}. Several observations can be made from the example: the performance of the building is slightly improved after the intelligent control system installation, as expected. However, during the heating season 2022-2023, something occurs that increases heating system consumption. We analyze this example further in Section~\ref{sec:changes} and show that this change is due to a ventilation-related operation unrelated to the heating control method. We conclude this section by stating that while such simple before-after comparisons can be used to track the \textit{overall} efficiency of the system, the method is not able to isolate the effect of the control system alone.

\subsection{Weather normalization methods}
\label{sec:normalization}

As weather conditions change from year to year, it is not possible to track the heating system performance accurately based only on raw consumption data. The goal of weather normalization methods is to eliminate the effect of weather by estimating what the consumption would have been under chosen reference conditions, denoted here by $T_{\mathrm{out}}^{\mathrm{ref}}$ and $\Phi_{\mathrm{rad}}^{\mathrm{ref}}$. Weather normalization thus attempts to make consumption data from different time periods comparable.

Most weather normalization approaches are \textit{ratio-based}, where power models are used to estimate the ratio between reference consumption and observed actual consumption. The normalization for total heating power then proceeds simply by
\begin{equation}
    \label{eq:norm_tot}
    q_{\mathrm{tot}}^{\mathrm{ref}} = \frac{f(T_{\mathrm{out}}^{\mathrm{ref}}, \Phi_{\mathrm{rad}}^{\mathrm{ref}})}{f(T_{\mathrm{out}}^{\mathrm{obs}}, \Phi_{\mathrm{rad}}^{\mathrm{obs}})} \cdot q_{\mathrm{tot}}^{\mathrm{obs}},
\end{equation}
where $f$ is a model for the total power consumption.

Some popular normalization approaches get a particularly simple form, if we only use the ratio-based normalization for the space heating part and treat DHW consumption separately:
\begin{equation}
    q_{\mathrm{tot}}^{\mathrm{ref}} = \frac{g(T_{\mathrm{out}}^{\mathrm{ref}}, \Phi_{\mathrm{rad}}^{\mathrm{ref}})}{g(T_{\mathrm{out}}^{\mathrm{obs}}, \Phi_{\mathrm{rad}}^{\mathrm{obs}})} \cdot q_{\mathrm{space}}^{\mathrm{obs}} + q_{\mathrm{dhw}},
\end{equation}
where $g$ is now the model for the space heating consumption.

However, we typically only have total power data, and the DHW contribution must therefore be estimated with $q_{\mathrm{dhw}}=q_{\mathrm{dhw}}^{\mathrm{est}}$, which yields the following widely used formula for normalization:
\begin{equation}
    \label{eq:norm_space}
    q_{\mathrm{tot}}^{\mathrm{ref}} = \frac{g(T_{\mathrm{out}}^{\mathrm{ref}}, \Phi_{\mathrm{rad}}^{\mathrm{ref}})}{g(T_{\mathrm{out}}^{\mathrm{obs}}, \Phi_{\mathrm{rad}}^{\mathrm{obs}})} \cdot (q_{\mathrm{tot}}^{\mathrm{obs}} - q_{\mathrm{dhw}}^{\mathrm{est}}) + q_{\mathrm{dhw}}^{\mathrm{est}}.
\end{equation}

This equation has a built-in problem though: when outdoor temperature increases, the space heating demand modeled by $g$ approaches zero. The division-by-zero issues and unstable results at warm outdoor conditions must then be circumvented with special heuristics. If the ratio is calculated for the total heat consumption, as in Eq.~\eqref{eq:norm_tot}, the problem disappears.

The models $f$ and $g$ naturally affect the accuracy of the normalization. A model similar to Eq.~\eqref{eq:gam} was tested in \cite{rapo25} within the weather normalization context, yielding promising and stable results while circumventing the issues mentioned for the model in Eq.~\eqref{eq:norm_space}. The drawback of this approach is that it requires relatively advanced statistical techniques for calibrating the model, such as those discussed in \cite{solonen23} and Appendix~\ref{sec:appendix_A}. 

A much simpler approach, without the need for any model calibration, is obtained by using Eq.~\eqref{eq:norm_space} and the simplest possible version of the power model \eqref{eq:gam} with just a linear outdoor temperature term:
\begin{equation}
    g(T_{\mathrm{out}}) =  b\left( T_{\mathrm{in}} - T_{\mathrm{out}}\right).
\end{equation}

Note that the indoor temperature dependency is neglected here, since that data is typically not available, and we need to assume a fixed value $T_{\mathrm{in}}=T_{\mathrm{base}}$. With this assumption, the building-specific constant $b$ cancels out and the normalization ratio becomes simple:
\begin{equation}
    q_{\mathrm{tot}}^{\mathrm{ref}} = \frac{T_{\mathrm{base}} - T_{\mathrm{out}}^{\mathrm{ref}}}{T_{\mathrm{base}} - T_{\mathrm{out}}^{\mathrm{obs}}} \cdot (q_{\mathrm{tot}}^{\mathrm{obs}} - q_{\mathrm{dhw}}^{\mathrm{est}}) + q_{\mathrm{dhw}}^{\mathrm{est}}.
\end{equation}

Typically, instead of normalizing hourly heating power readings, we want to normalize the observed energy consumption $E_{\mathrm{tot}}^{\mathrm{obs}} = \sum{q_{\mathrm{tot}}^{\mathrm{obs}}}$ over longer time periods (e.g., monthly). In this case, the normalization requires applying appropriate sums over the desired time periods:
\begin{equation}
    E_{\mathrm{tot}}^{\mathrm{ref}} = \frac{\sum (T_{\mathrm{base}} - T_{\mathrm{out}}^{\mathrm{ref}})}{\sum (T_{\mathrm{base}} - T_{\mathrm{out}}^{\mathrm{obs}})} \cdot (E_{\mathrm{tot}}^{\mathrm{obs}} - E_{\mathrm{dhw}}^{\mathrm{est}}) + E_{\mathrm{dhw}}^{\mathrm{est}},
\end{equation}
where $E_{\mathrm{dhw}}^{\mathrm{est}}=\sum q_{\mathrm{dhw}}^{\mathrm{est}}$ is the estimated DHW energy consumption over the desired period. This gives rise to the concept of \textit{Heating Degree Days} (HDDs) over the period, and the normalization formula finally reads as
\begin{equation}
    E_{\mathrm{tot}}^{\mathrm{ref}} = \frac{\mathrm{HDD}_{\mathrm{ref}}}{\mathrm{HDD}_{\mathrm{obs}}} \cdot (E_{\mathrm{tot}}^{\mathrm{obs}} - E_{\mathrm{dhw}}^{\mathrm{est}}) + E_{\mathrm{dhw}}^{\mathrm{est}},
\end{equation}
where $\mathrm{HDD}_{\mathrm{ref}} = \sum (T_{\mathrm{base}} - T_{\mathrm{out}}^{\mathrm{ref}})$ are the heating degree days of the reference year and $\mathrm{HDD}_{\mathrm{obs}} = \sum (T_{\mathrm{base}} - T_{\mathrm{out}}^{\mathrm{obs}})$ are the observed HDDs at the building location. HDDs depend only on weather (assuming a fixed $T_{\mathrm{base}}$), and the HDDs for the reference year and current year are thus often available directly from national meteorological institutes\footnote{For instance, the HDD values for Finland can be obtained from \url{https://en.ilmatieteenlaitos.fi/heating-degree-days}}. This yields a simple normalization recipe, once we decide on a few practical matters: (i) what is the chosen $T_{\mathrm{base}}$, (ii) how do we estimate $E_{\mathrm{dhw}}^{\mathrm{est}}$, and (iii) how do we handle low space heating consumption periods (e.g., spring) to avoid unstable results and division-by-zero issues. Detailed discussion of these choices is beyond the scope of the paper, but we provide an example of some procedures often applied in Finland (as recommended by the Finnish government owned company \textit{Motiva}) in Appendix~\ref{sec:appendix_B}.

\begin{figure}[!ht]
    \centering
    \includegraphics[width=0.85\linewidth]{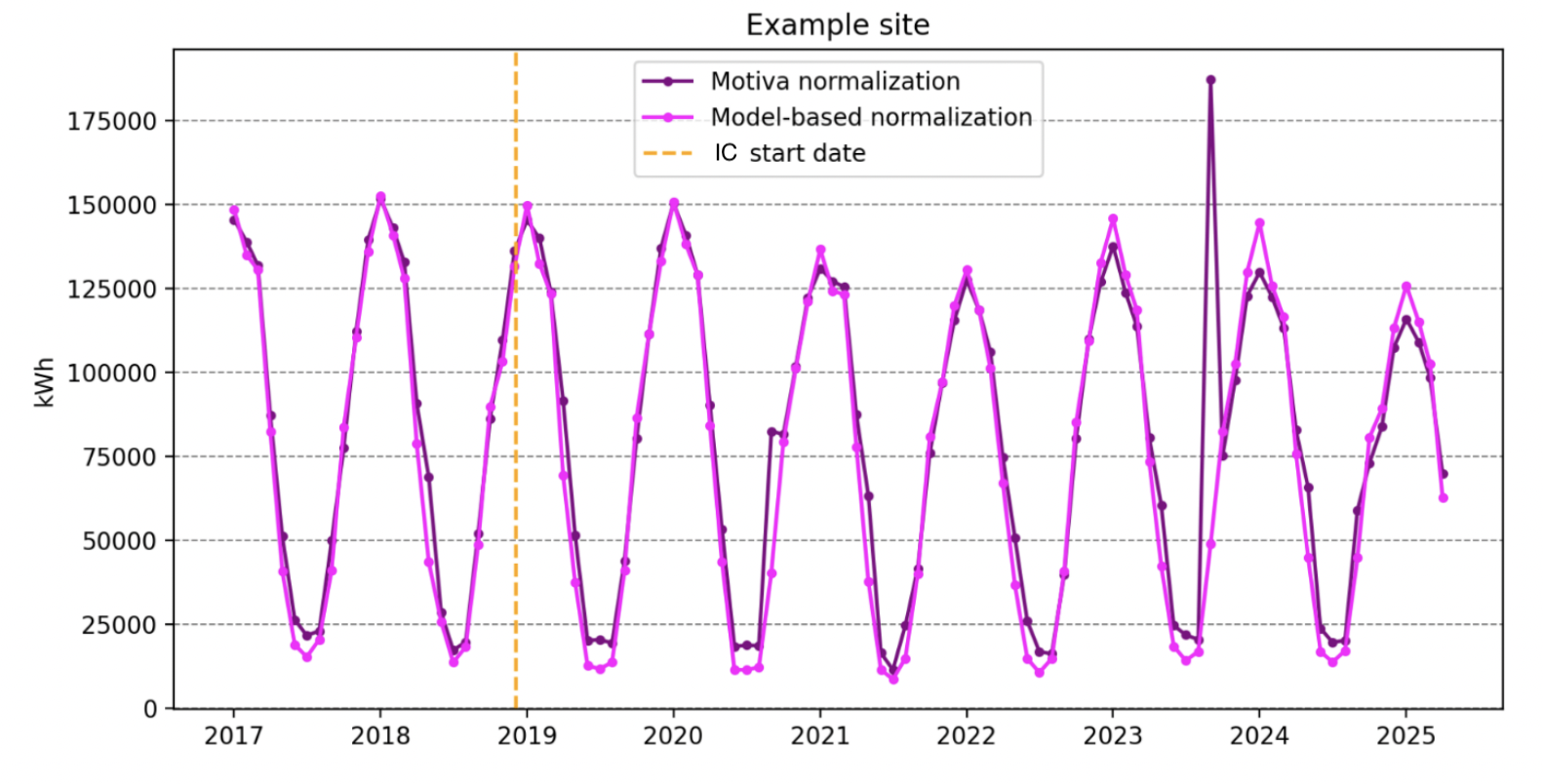}
    \caption{Comparison of Motiva normalization and more advanced model-based normalization for an example building. Reproduced with author's permission from \cite{rapo25}.}
    \label{fig:weather_norm}
\end{figure}

The HDD-based weather normalization approach derived above is likely the most widely used building performance tracking method. It is often described as a ``model-free alternative'', but, as the above derivation shows, a very crude power model is involved in the background. A comparison between HDD-based normalization and the same ratio-based approach with a more sophisticated power model is given in Fig.~\ref{fig:weather_norm}. We observe that improving the normalization model accuracy and applying the ratio-based normalization to total power instead of space heating only (Eq.~\eqref{eq:norm_tot}) produces more stable results, especially in the spring and autumn when HDD-based methods encounter problems. Moreover, a more sophisticated model can incorporate the contribution of solar radiation into the normalization, whereas the HDD-based methods typically account only for outdoor temperature.

In summary, weather normalization can yield a good and simple approach for building performance tracking. However, the method still targets the overall performance of the building, and is unable to isolate the effects of intelligent control.

\subsection{Supply temperature based methods}
\label{sec:supplytemp}

An alternative approach for evaluating intelligent control effects, distinct from the power model-based methods discussed above, employs supply temperature based comparisons. Often, we have a good understanding of the heating curve used before installing the intelligent control system, and we can thus simulate what the supply temperatures would have been under standard heating curve control. The task then is to map the supply temperature difference between the heating curve simulation and the observed values to power consumption, which can be done using the models discussed in Section~\ref{sec:supplytemp}. Note that this method targets control effects only, not the total heat consumption changes.

More formally, let us denote by $T_{\mathrm{sup}}^{\mathrm{HC}}$ the supply temperature given by the heating curve (as a function of outdoor temperature). Moreover, let $T_{\mathrm{sup}}^{\mathrm{obs}}$ denote the observed supply temperature values under the intelligent control period. We can then calculate the relative difference in heating power by applying the supply temperature vs. power model in  Eq.~\eqref{eq:pow_vs_sup}:
\begin{equation}
    \mathrm{effect} = \frac{q_{\mathrm{tot}}^{\mathrm{obs}}-q_{\mathrm{tot}}^{\mathrm{HC}}}{q_{\mathrm{tot}}^{\mathrm{HC}}} = \frac{k \left( T_{\mathrm{sup}}^{\mathrm{obs}}-T_{\mathrm{in}} \right) + q_{\mathrm{dhw}}-k \left( T_{\mathrm{sup}}^{\mathrm{HC}}-T_{\mathrm{in}} \right) - q_{\mathrm{dhw}}}{k \left( T_{\mathrm{sup}}^{\mathrm{HC}}-T_{\mathrm{in}} \right) + q_{\mathrm{dhw}}} = \frac{k(T_{\mathrm{sup}}^{\mathrm{obs}}-T_{\mathrm{sup}}^{\mathrm{HC}})}{k \left( T_{\mathrm{sup}}^{\mathrm{HC}}-T_{\mathrm{in}} \right) + q_{\mathrm{dhw}}}.
\end{equation}

The calculation would then proceed by calibrating the coefficients $k$ and $q_{\mathrm{dhw}}$ from data. Note that in the above equation we have assumed that the indoor temperature and DHW consumption are the same under both heating curve control and intelligent control. DHW can be assumed to remain unchanged, since we do not control the DHW circuit. However, indoor temperature typically changes a slightly when the control model is changed. This introduces a small bias in the results, although the effect is quite small. 
An additional assumption here is that the flow rate in the heat circulation network stays unchanged, which might not be realistic in all cases.

An even simpler formula can be derived by calculating relative differences for the space heating part alone:
\begin{equation}
    \mathrm{effect} = \frac{q_{\mathrm{space}}^{\mathrm{obs}}-q_{\mathrm{space}}^{\mathrm{HC}}}{q_{\mathrm{space}}^{\mathrm{HC}}} = \frac{k \left( T_{\mathrm{sup}}^{\mathrm{obs}}-T_{\mathrm{in}} \right) - k \left( T_{\mathrm{sup}}^{\mathrm{HC}}-T_{\mathrm{in}} \right) }{k \left( T_{\mathrm{sup}}^{\mathrm{HC}}-T_{\mathrm{in}} \right)} = \frac{T_{\mathrm{sup}}^{\mathrm{obs}}-T_{\mathrm{sup}}^{\mathrm{HC}}}{ T_{\mathrm{sup}}^{\mathrm{HC}}-T_{\mathrm{in}}}.
\end{equation}

That is, we can get an estimate of the \textit{space heating effect} by just comparing supply temperatures, and the building-specific constants $k$ cancel out. This is the only method that works even without having power data available.

Supply temperature based methods are simple to implement and target only the control effects, which is what we want. However, the method has a few caveats compared to the other approaches. First, we might not always have the correct original heating curve available. There are data-based ways to estimate what kind of heating curve was used before the intelligent control system, but that is outside the scope of this paper. Moreover, the method assumes that the power is linear in supply temperature, which assumes that we are in the supply temperature control regime, as discussed in Section $\ref{sec:te1_vs_pow}$. From the data, we see that with the heating curve we are often in this regime in the Nordic countries, but we also see some thermostat effects, especially when the solar radiation intensity is high. Capturing these thermostat effects in the models is challenging, and \textit{we therefore recommend using these supply temperature-based methods as the ``last resort'' when no other method is available} due to, e.g., data limitations.



\section{Efficiency changes in buildings}
\label{sec:changes}

The main challenge in evaluating the consumption effects of intelligent control solutions is that building performance keeps changing all the time due to various reasons not related to heating control. Only the supply temperature based methods discussed above can be used to gain insight into control effects alone, but they have their own issues, as discussed in Section \ref{sec:supplytemp}. In this section, we discuss different non-control-related performance changes and show how they can be categorized and studied based on data.

Building consumption level changes that are not due to control method can be roughly divided into 5 categories:
\begin{enumerate}
    \item Slow changes due to aging of the building structures and heating equipment. In the literature, this effect is estimated to be as high as 20-30\% during a 20-year period, see \cite{eleftheriadis17}. The magnitude of these slow changes naturally depend on various maintenance operations done to the building.
    \item Changes in the ventilation system. If the ventilation power is increased, the building consumption likely also increases as more heat escapes outside the building.
    \item Changes in the heating system such as hydronic balancing, equipment change and maintenance, circulation pump operations, etc.
    \item Large renovations done at the buildings.
    \item Changes in occupancy and domestic hot water usage (see, e.g., \cite{sun2020review} for a review of occupancy-driven energy variations).
\end{enumerate}

\begin{figure}[!ht]
    \centering
    \includegraphics[width=1.0\linewidth]{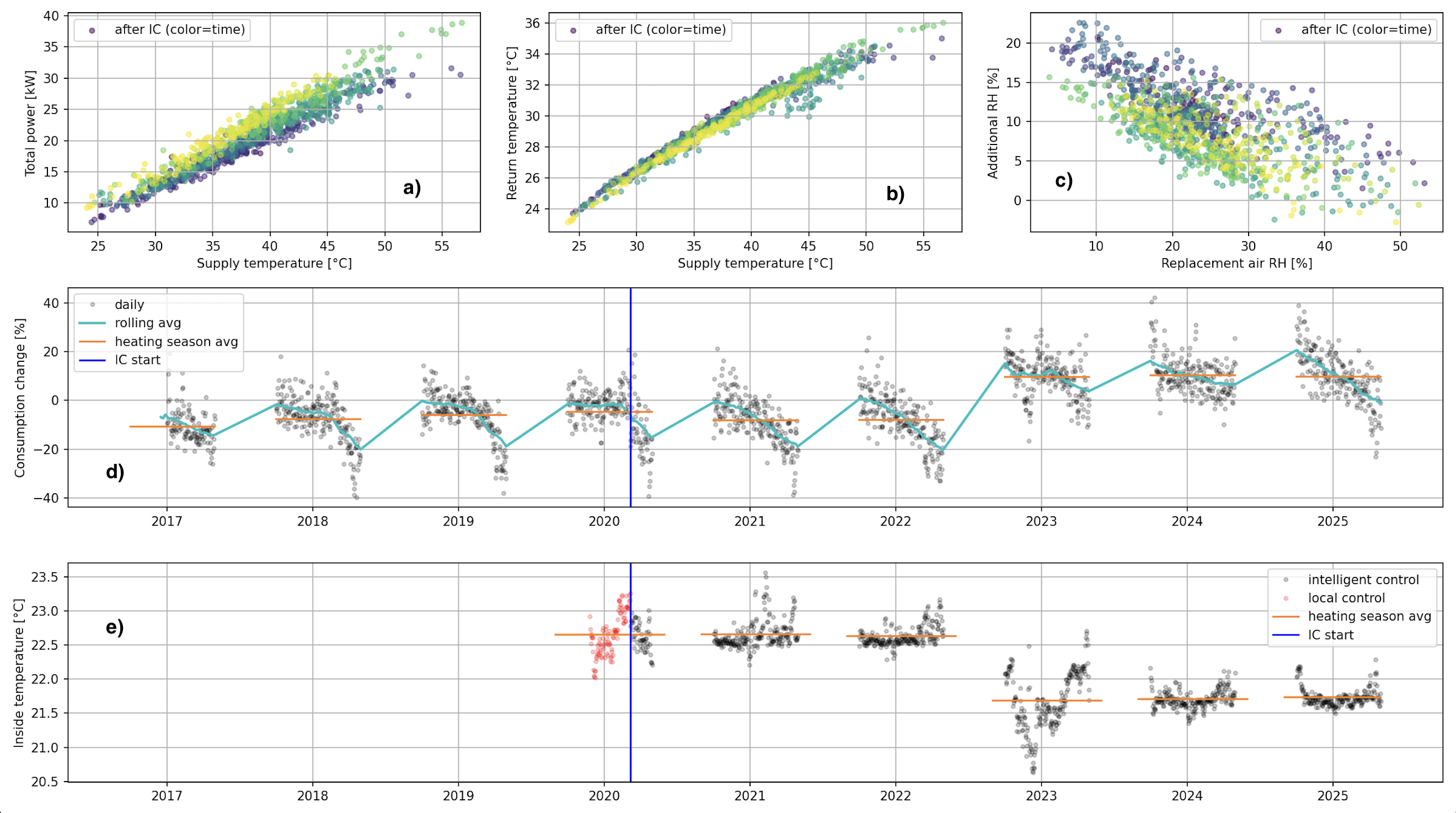}
    \caption{Performance change analysis for an example building. Subplots: a) total power vs. supply temperature, b) return temperature vs. supply temperature, c) additional relative humidity vs. replacement air relative humidity, d) consumption change percentage and e) indoor temperature. Vertical lines indicate the intelligent control start date.}
    \label{fig:changes}
\end{figure}

Good visibility into these performance changes can be gained by examining the example in Fig.~\ref{fig:changes}. The figure is divided into five subplots that each illustrate different aspects of building behavior. Color in the scatter plots indicate time. The figure consists of the following components:
\begin{itemize}
    \item[a)] Total power consumption vs. supply temperature relationship is useful for determining whether changes have occured at the heating system level (category 3 above). For example, the same supply temperature suddenly yields higher power consumption. The figure is also useful in detecting changes in domestic hot water consumption (category 5): in that case, the intercept of the supply temperature vs. power relationship would change (higher power consumption at all supply temperatures). Various heating system changes (e.g. hydronic balancing) can affect the slope of the relationship. In this example, the relationship stays rather constant during the time of the consumption change, indicating no sudden changes at the heating system level.
    \item[b)] Supply temperature vs. return temperature relationship can reveal changes in the circulation of the heating network (category 3). For instance, pump setting changes or actions at the thermostat level (e.g., balancing) would be visible in this graph. In this example, the relationship stays constant, indicating no significant changes in the circulation either.
    \item[c)] Replacement air relative humidity vs. additional relative humidity\footnote{Replacement air relative humidity (RH) is a "computational metric" which represents the RH when replacement air from the outdoors is heated to the temperature of the indoor air. It is calculated based on outdoor temperature, outdoor RH, and indoor temperature. Additional relative humidity is the difference between measured indoor RH and the replacement air RH.} relationship can indicate changes done to the ventilation system (category 2): increasing ventilation power results in lower additional humidity inside the apartments. Humidity data is often measured by the same sensors as indoor temperatures in intelligent control systems. In the example here, there is a change to lower indoor humidity levels at one point in time.
    \item[d)] Measured total heat consumption vs. modeled consumption (Eq.~\eqref{eq:modelbased}) gives a weather-normalized ``performance time series'' of the building, from which we can detect major changes in consumption. The power model formulation can be, e.g., one of the versions discussed in Section \ref{sec:modelbased}. Here, the model is calibrated using data from the intelligent control period, and higher values mean higher consumption. In this example, the consumption increases significantly at the same time humidity levels decreased. This indicates that a ventilation-related operational change at the building is likely responsible.
    \item[e)] Indoor temperature time series indicates whether a consumption change is related to indoor temperature level changes (control-related) or something else. In this example, the indoor temperatures have been lowered around the time the consumption increase occurs, whereas lower indoor temperatures should correspond to lower consumption. Isolating the control-related effects from the ventilation change would enable us to quantify this.
\end{itemize}

In summary, buildings constantly undergo changes that are not related to control, like the ventilation-related change discussed above. More examples of the various change categories are given in Appendix~\ref{sec:appendix_D}. These changes present a challenge for all discussed performance tracking approaches that target the total heat consumption: how to isolate control-related consumption changes from other changes? Our approach to this is discussed next. 


\section{Isolating the intelligent control effects}
\label{sec:lheffect}

\begin{figure}[!ht]
    \centering
    \includegraphics[width=1.0\linewidth]{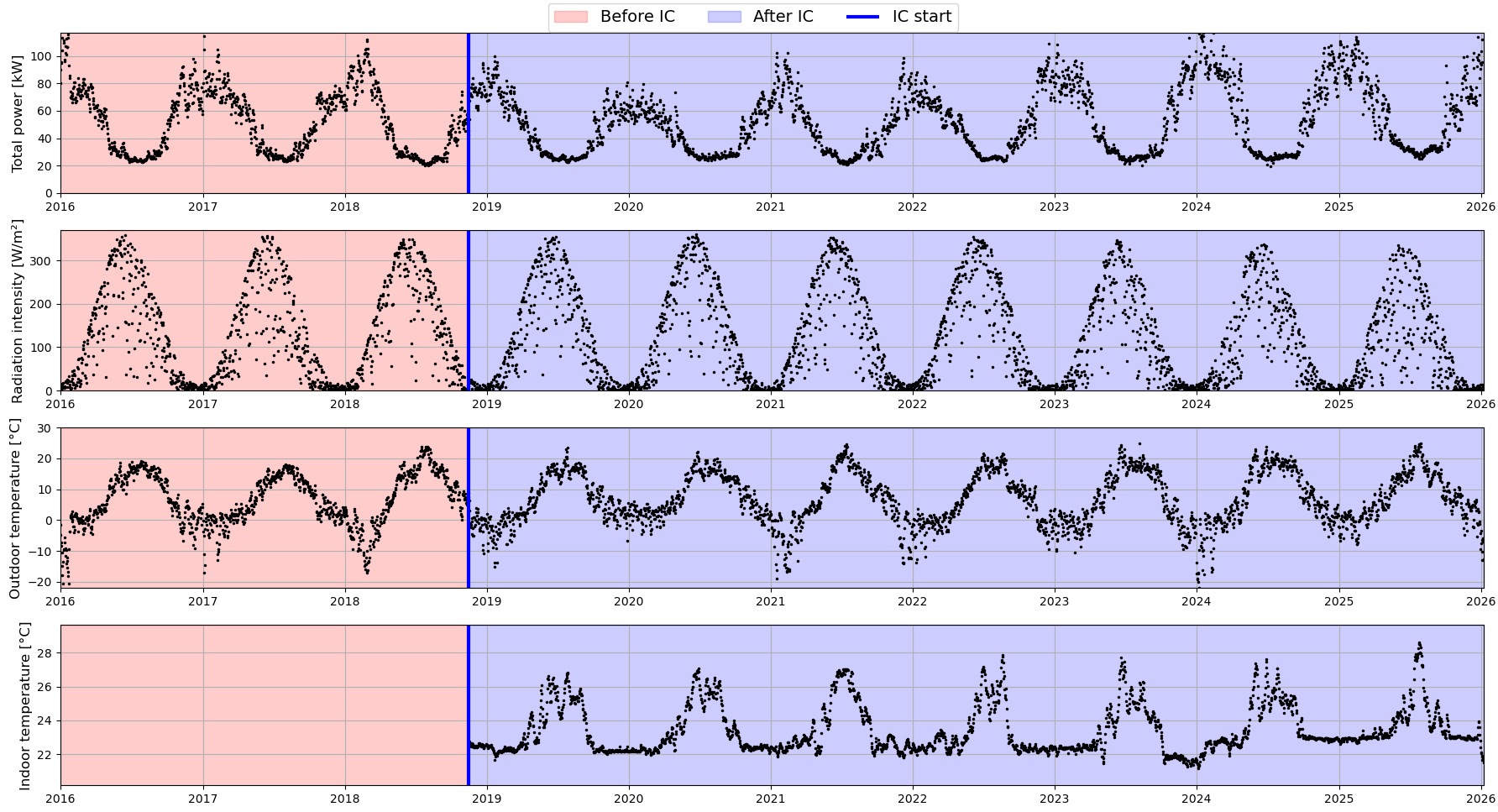}
    \caption{Data periods for the model-based before-after comparison.}
    \label{fig:periods_old}
\end{figure}

Our method for isolating control effects from other performance changes builds upon building two consumption models, similarly as in approach 4 of \cite{saloux25}: one that characterizes the behavior before intelligent control activation and another one for the period after the activation. The key here is the selection of the model calibration periods. Typically, in model-based performance tracking methods discussed in Section~\ref{sec:modelbased}, all data before and after an event is taken into account, as depicted in Fig.~\ref{fig:periods_old}. The challenge here is that buildings undergo non-control-related performance changes, as discussed in Section~\ref{sec:changes}, which affects the model calibration and results. Moreover, the applied models usually consider only weather variables (outside temperature, solar radiation) as inputs, and ignore control-related parameters, most notably the indoor temperature changes and the targeted indoor temperature levels.

To make the approach less vulnerable to performance changes, we consider only data right before and after the intelligent control activation, typically meaning roughly one heating season worth of data before and after the event. We assume that the intelligent control activation is the only major consumption-related event in the building. It is obviously possible that other events occur in the model calibration periods also, but at least this is an improvement compared to assuming that no other changes ever happen.

\begin{figure}[!ht]
    \centering
    \includegraphics[width=1.0\linewidth]{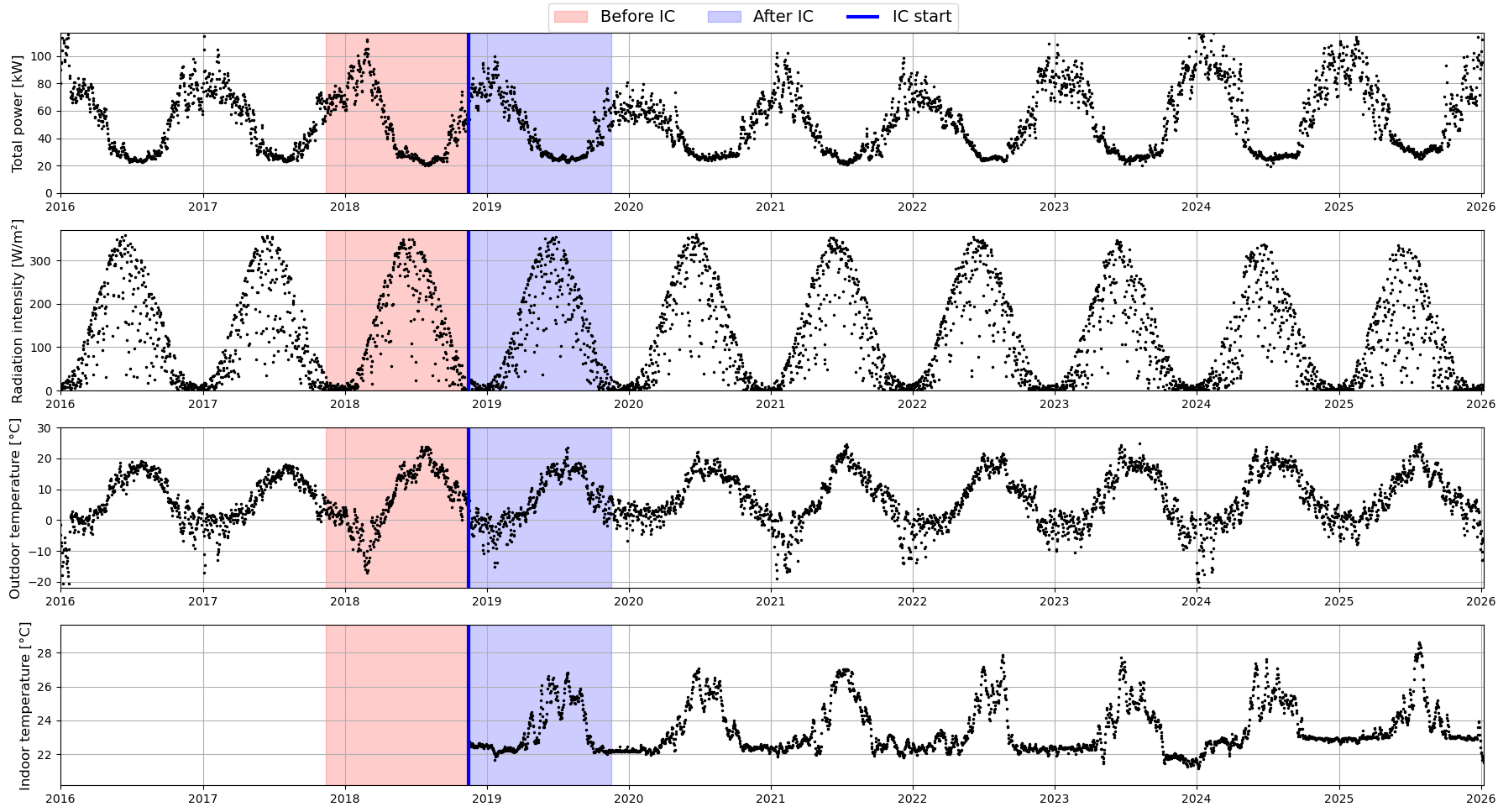}
    \caption{Data periods for the intelligent control effect calculation.}
    \label{fig:periods_new}
\end{figure}

With the two models, we are then able to extrapolate (simulate) the control effects beyond the considered model calibration periods, as pointed out in \cite{saloux25}. However, during the intelligent heating control the indoor temperature levels can be significantly changed by the end user, which affects the consumption, and it is thus important that the models contain indoor temperature as their input. Our solution is to employ a gray-box model similar to the ones in Eq.~\eqref{eq:gam}, which fulfill this requirement. Note, however, that indoor temperature data is typically only available after the intelligent control installation, so the model for the pre-installation period needs to calibrate some average unknown indoor temperature level. When the previous control approach was a rule-based heating curve control method, this is well justified, as no specific indoor temperature is targeted with such control, and only outside temperature and solar radiation intensity (through the thermostat effect) are the determinants of the heating power of such heating method. The input data and calibration periods for our approach are illustrated in Fig.~\ref{fig:periods_new}.

To express the idea more precisely, let us denote by $m_{\mathrm{ref}}(T_{\mathrm{out}}, \Phi_{\mathrm{rad}})$ the model for the period before the intelligent control activation, and by $m_{\mathrm{ic}}(T_{\mathrm{out}}, \Phi_{\mathrm{rad}}, T_{\mathrm{in}})$ the model after the activation. Note that in the reference model there is no dependency on indoor temperature since we typically have no data for it prior to intelligent control activation. We thus assume that the reference model describes the consumption behavior with some average indoor temperature level realized with the old control method. The models applied here are of the form
\begin{subequations}
\begin{align}
    m_{\mathrm{ref}}(T_{\mathrm{out}}, \Phi_{\mathrm{rad}}) &=  f_1\left( T_{\mathrm{base}} - T_{\mathrm{out}}\right) - f_2(T_{\mathrm{out}})\Phi_{\mathrm{rad}} + q_{0,\mathrm{ref}}  \\
    m_{\mathrm{ic}}(T_{\mathrm{out}}, \Phi_{\mathrm{rad}}, T_{\mathrm{in}}) &= g_1\left( T_{\mathrm{in}} - T_{\mathrm{out}}\right) - g_2(T_{\mathrm{out}})\Phi_{\mathrm{rad}} + q_{0,\mathrm{ic}}.
\end{align}
\end{subequations}

Note that, compared to Eq.~\eqref{eq:gam}, the indoor temperature change term $C_{\mathrm{in}} \nicefrac{\dd T_{\mathrm{in}}}{\dd t}$ is neglected here. As discussed in \cite{rapo25}, this effect is small compared to the other effects in the model. Moreover, during intelligent control, indoor temperatures usually are kept around a given setpoint, so the indoor temperature change effect can be assumed to be negligible, especially when considering that we are looking at performance at daily resolution.

Choosing $T_{\mathrm{base}}$ might seem difficult, since the indoor temperature is unknown before the system installation. However, its value does not affect the model estimates: the error in $T_{\mathrm{base}}$ (which can be outdoor temperature dependent) is indirectly accounted for in the model calibration. That is, we can safely assume a fixed value for $T_{\mathrm{base}}$ here. 

We remark that the functions in the previous equations are generally nonlinear. We apply Bayesian GAMs to parameterize them. See \cite{solonen23} for some Bayesian GAM theory and Appendix \ref{sec:appendix_A} for details of our GAM implementation.

After the models have been calibrated, we can define the intelligent control effect in power units as
\begin{equation}
    \mathrm{effect}_{\mathrm{ic}}\left( T_{\mathrm{out}}, \Phi_{\mathrm{rad}}, T_{\mathrm{in}} \right) = m_{\mathrm{ic}}(T_{\mathrm{out}}, \Phi_{\mathrm{rad}}, T_{\mathrm{in}}) - m_{\mathrm{ref}}(T_{\mathrm{out}}, \Phi_{\mathrm{rad}}).
\end{equation}

That is, the effect now depends only on weather inputs and the chosen indoor temperature level $T_{\mathrm{in}}$ chosen by the end user. Now we can subtract this intelligent control effect from the observed power readings to get an estimate of the power consumption without intelligent control:
\begin{subequations}
\label{eq:lh_old_effect}
\begin{align}
    q_{\mathrm{ic}} &=  q_{\mathrm{tot}}^{\mathrm{obs}}  \\
    q_{\mathrm{old}} &= q_{\mathrm{tot}}^{\mathrm{obs}} - \mathrm{effect}_{\mathrm{ic}}.
\end{align}
\end{subequations}

\begin{figure}[!ht]
    \centering
    \includegraphics[width=0.8\linewidth]{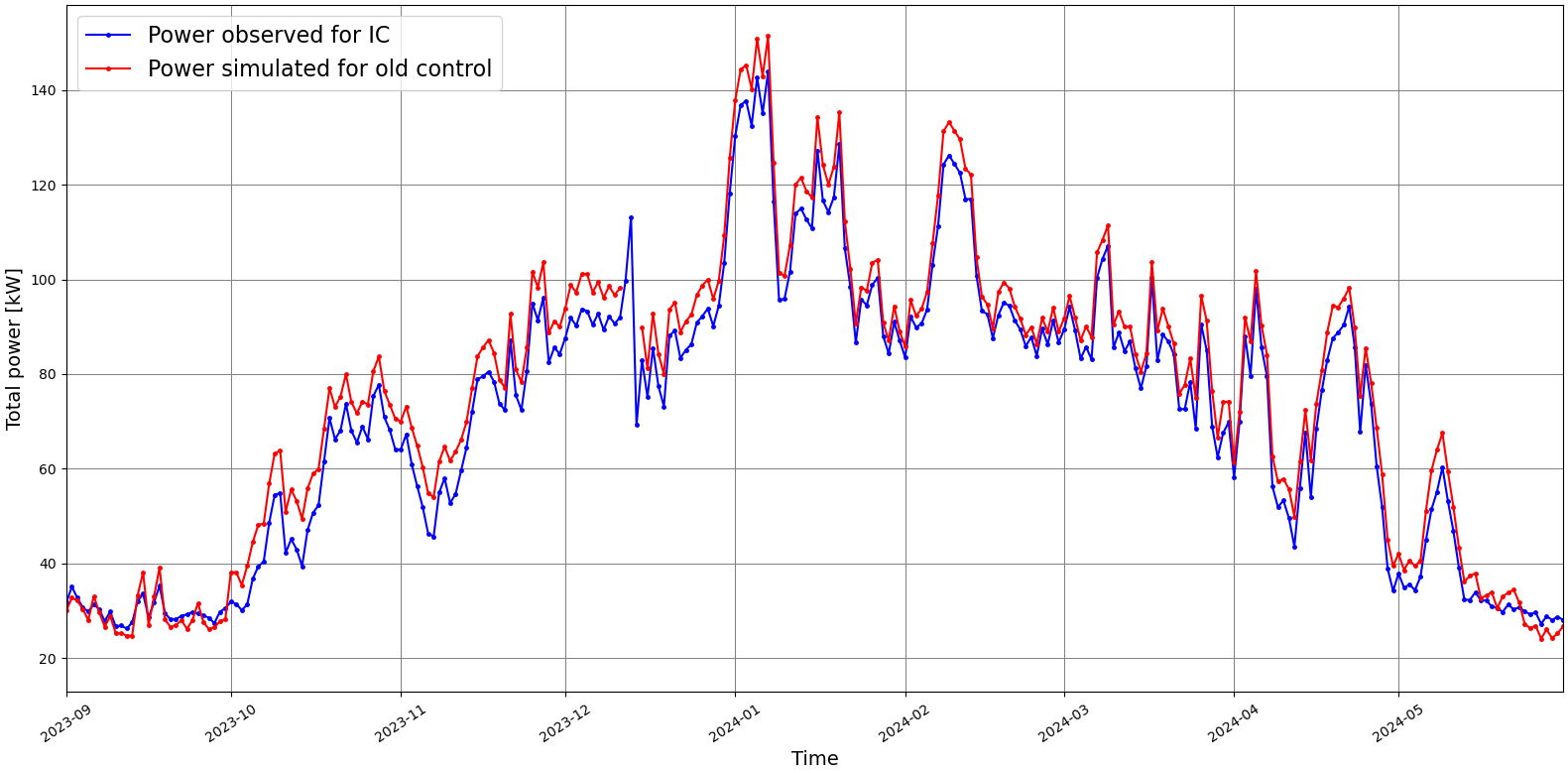}
    \caption{Example of measured daily average power for a heating season (blue) and simulated consumption with the old control system (red).}
    \label{fig:power_comp}
\end{figure}

The relative comparison between the new and old method can now be done based on Eqs.\eqref{eq:lh_old_effect}. An example of power consumption with intelligent control (observed power) and an estimate of consumption with the old control system is given in Fig.~\ref{fig:power_comp}.

\begin{figure}[!ht]
    \centering
    \includegraphics[width=0.95\linewidth]{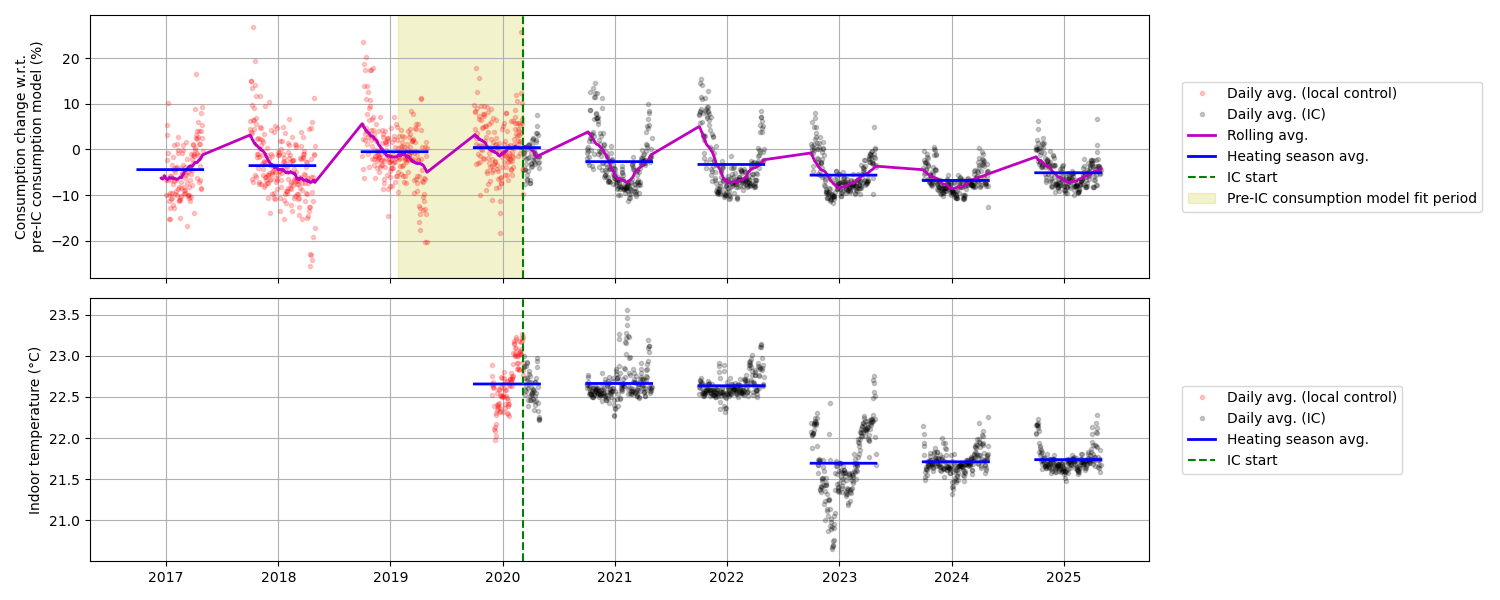}
    \caption{Relative comparison between the estimated consumption with the old control and intelligent control (top) and realized indoor temperatures (bottom).}
    \label{fig:perf_change_new}
\end{figure}

Now, the intelligent control effect is simulated with the two models into the future conditions, and depends only on weather and indoor temperature. That is, this method no longer reacts to non-control related performance changes. As an example, let us take another look at the ventilation example discussed in Fig.~\ref{fig:changes} in Section~\ref{sec:changes}. The relative comparison between consumption with the old and new control is given in Fig. \ref{fig:perf_change_new}. We observe that the consumption change due to ventilation is no longer visible, and lowering the indoor temperature reduces the consumption, as expected.

\subsection{Decomposing the intelligent control effect}

With the presented simulation-based approach, we can go a step further from evaluating the total effect, and decompose the control effect into sub-components by simulating the models in different ways. This type of analytics can answer interesting questions like \textit{what part of the effect came from solar radiation?} or \textit{what was the effect of the introduced night and weekend setbacks?}.

To answer these questions, we can use the models to simulate the intelligent control effect \textit{without} solar radiation or setbacks by modifying the input data. For instance, the solar radiation effect can be calculated by simulating the effect with solar radiation set to zero and subtracting the result from the total effect:
\begin{equation}
    \mathrm{effect}_{\mathrm{rad}} = \mathrm{effect}_{\mathrm{ic}}\left( T_{\mathrm{out}}, \Phi_{\mathrm{rad}}, T_{\mathrm{in}} \right) - \mathrm{effect}_{\mathrm{ic}}\left( T_{\mathrm{out}}, 0, T_{\mathrm{in}} \right).
\end{equation}

Similarly, the setback effect can be estimated by defining the indoor temperature without setbacks as $T_{\mathrm{in}}^{(\mathrm{no\ setbacks})}$ and calculating 
\begin{equation}
    \mathrm{effect}_{\mathrm{setback}} = \mathrm{effect}_{\mathrm{ic}}\left( T_{\mathrm{out}}, \Phi_{\mathrm{rad}}, T_{\mathrm{in}} \right) - \mathrm{effect}_{\mathrm{ic}}\left( T_{\mathrm{out}}, \Phi_{\mathrm{rad}}, T_{\mathrm{in}}^{(\mathrm{no\ setbacks})} \right).
\end{equation}

For other control-related components, the calculation proceeds similarly: individual effects can be calculated by dropping the effects from the calculation one-by-one. Finally, to ``close the loop'' and ensure the sum of the sub-components match the total effect, we can define an effect called \textit{other}, which includes all control effects not captured in the sub-components. In this example, after calculating the solar radiation and setback effects, the other effect would be
\begin{equation}
    \mathrm{effect}_{\mathrm{other}} = \mathrm{effect}_{\mathrm{ic}} - \mathrm{effect}_{\mathrm{rad}} - \mathrm{effect}_{\mathrm{setback}}.
\end{equation}

In the example here, the other-effect would include, for instance, the effects of changing the indoor temperature setpoints.

\begin{figure}[!ht]
    \centering
    \includegraphics[width=1.0\linewidth]{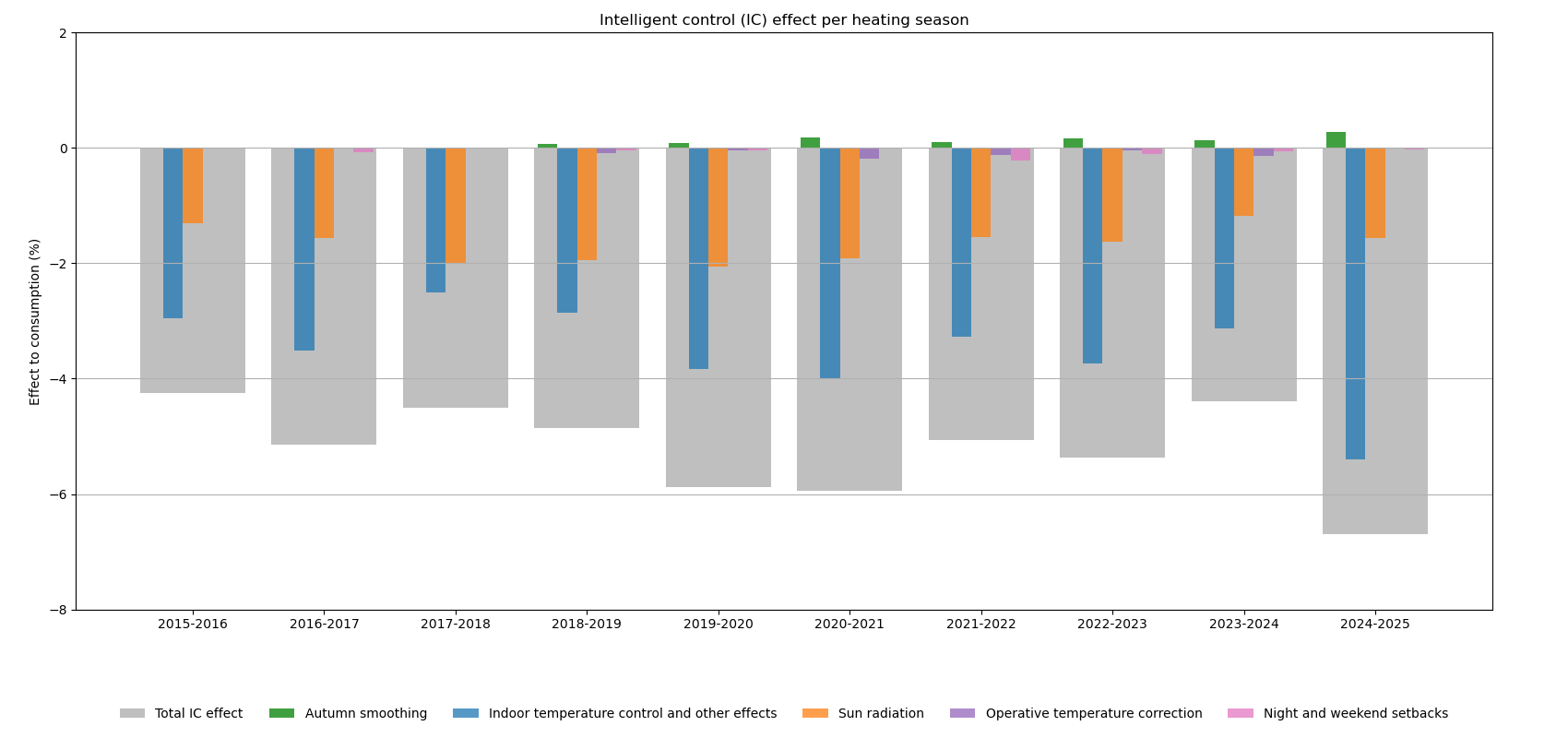}
    \caption{Example of an intelligent control effect decomposition for an example building over different heating seasons.}
    \label{fig:decomp_demo}
\end{figure}

An example of such analytics over heating seasons averaged for one building portfolio can be seen in Fig.~\ref{fig:decomp_demo}. We observe, for example, that different years have varying solar radiation contributions (e.g., season 2019-2020 was exceptionally sunny) and that the improved savings results over the years are mostly due to the other-component (lowered indoor temperatures). This kind of analytics can provide valuable insight into the effects of various control-related decisions made by the end user. We remark that here we have more sub-components than just the solar radiation and setbacks mentioned above, such as ``operative temperature correction'' and ``autumn smoothing''. These are specific control features and are out of scope here.

\section{Discussion and conclusions}
\label{sec:conclusions}

In this paper, we reviewed and derived several popular performance evaluation methods, including weather normalization and model-based approaches. We discussed how the existing methods track overall performance and thus fail to isolate control-related consumption effects from other changes. These changes were categorized and analyzed with real-life examples.

As the main novel contribution, we presented a simulation-based technique for isolating and tracking exclusively control-related consumption effects. Furthermore, we demonstrated how this model-based approach can be used to decompose the effect into sub-components, such as solar radiation and setback effects. The methodology was validated using an example from a real building.

While the proposed method offers improvements in savings verification, it is not without limitations. Its accuracy relies on the availability of ``clean'' calibration periods before and after control system installation, with the assumption that no other major changes overlap with these windows. Future implementations could benefit from automated anomaly detection, based on the analytics discussed in Section \ref{sec:changes}, to ensure the validity of the calibration periods.

In addition to energy savings, intelligent control approaches often have other benefits, such as peak power shaving via load shifting. The evaluation of these benefits presents its own challenges, which are left for future work.

\appendix
\section{Calibrating GAM power models}
\label{sec:appendix_A}

Consider the following GAM formulation
\begin{equation}
    y_i = f_1(\mathbf{x}_{i,1}, \pmb{\theta}_1)+f_2(\mathbf{x}_{i,2}, \pmb{\theta}_2)+\cdots+f_M(\mathbf{x}_{i,M},\pmb{\theta}_M)+\varepsilon_i,
\end{equation}
where $f_i$ are unknown functions, $y_i$ is observation, $\mathbf{x}_{i,j}$ are input observations needed in function $j$, $\pmb{\theta}_i$ are some unknown parameters for function $f_i$ and $\varepsilon_i$ are measurement errors.

In GAMs, instead of using a parametric form for the unknown functions $f_i$, we wish to estimate them using non-parametric techniques. For that purpose, in most cases, we are able to write the functions using a linear combination of chosen basis functions:
\begin{equation}
    (f_i)_j = \sum_{k=1}^{K_i} \alpha_{i,k} p_{i,k}(\mathbf{x}_{i,j}),
\end{equation}
where $p$ are the chosen basis functions, and the unknown parameters related to the function are the weights, $\pmb{\theta}_i=[\alpha_1, ..., \alpha_K]$. Moreover, we often can assume something about the behavior of the functions (e.g., smoothness). This \textit{prior knowledge} can be included in the estimation using Bayesian methods, see \cite{gelman13} for a general introduction. Assuming that the measurement errors are Gaussian distributed and applying Gaussian distributed priors, our Bayesian GAM model can be written as
\begin{subequations}
\begin{align}
\label{pos1}
\mathbf{y} &=  \mathbf{A}\pmb{\theta} + \pmb{\varepsilon} \\ 
\pmb{B\theta} &\sim \mathcal{N} \left( \pmb{\mu}_{\mathrm{pr}}, \pmb{\Gamma}_{\mathrm{pr}} \right),
\label{post}
\end{align}
\end{subequations}
where $\mathbf{y}=[y_1, \cdots, y_N]^T$ is the vector of all observations, and the system matrix $\mathbf{A}$ takes the stacked form
\begin{equation}
    \mathbf{A} = 
    \begin{bmatrix}
        \overbrace{p_{1,1}(\mathbf{x}_{1,1})\ \ \cdots\ \ p_{1,K_1}(\mathbf{x}_{1,1})}^{f_1} & \overbrace{p_{2,1}(\mathbf{x}_{2,1})\ \ \cdots\ \ p_{2,K_2}(\mathbf{x}_{2,1})}^{f_2} & \cdots & \overbrace{p_{M,1}(\mathbf{x}_{M,1})\ \ \cdots\ \ p_{M,K_M}(\mathbf{x}_{M,1})}^{f_M} \\
        
        p_{1,1}(\mathbf{x}_{1,2})\ \ \cdots\ \ p_{1,K_1}(\mathbf{x}_{1,2}) & p_{2,1}(\mathbf{x}_{2,2})\ \ \cdots\ \ p_{2,K_2}(\mathbf{x}_{2,2}) & \cdots & p_{M,1}(\mathbf{x}_{M,2})\ \ \cdots\ \ p_{M,K_M}(\mathbf{x}_{M,2}) \\
        
        \vdots & \vdots & \vdots & \vdots \\
        
        p_{1,1}(\mathbf{x}_{1,N})\ \ \cdots\ \ p_{1,K_1}(\mathbf{x}_{1,N}) & p_{2,1}(\mathbf{x}_{2,N})\ \ \cdots\ \ p_{2,K_2}(\mathbf{x}_{2,N}) & \cdots & p_{M,1}(\mathbf{x}_{M,N})\ \ \cdots\ \ p_{M,K_M}(\mathbf{x}_{M,N}) \\
        
        & & & 
    \end{bmatrix}
\end{equation}

The posterior distribution of this model is available analytically, see \cite{solonen23} for the exact formulas in this case. A point estimate of the resulting posterior distribution, typically the posterior mean, is then used to represent the solution of the model calibration.

\begin{figure}[!ht]
    \centering
    \includegraphics[width=0.7\linewidth]{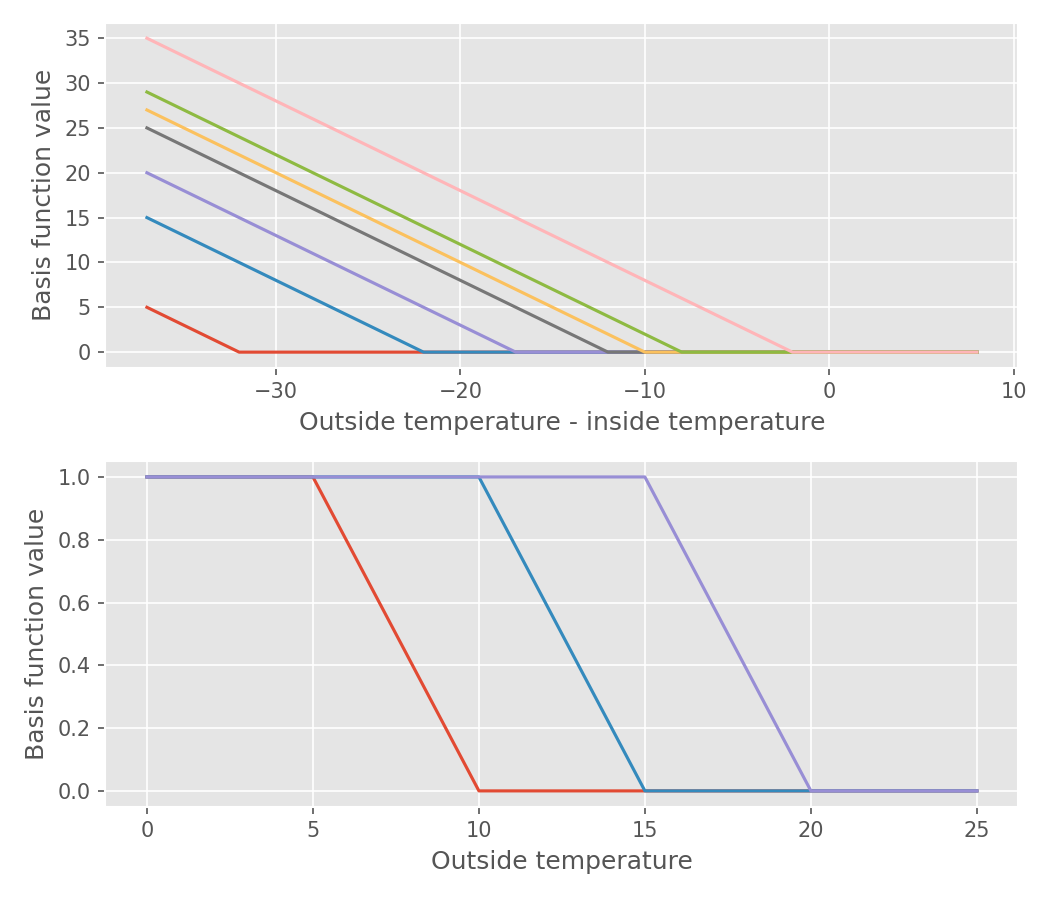}
    \caption{Chosen basis functions for $f_1$ (top) and $f_2$ (bottom). For $f_1$, a finer grid around the region where heating demand starts is chosen.}
    \label{fig:gammy_basis}
\end{figure}

In this paper, we wish to apply GAMs to power models of the form
\begin{equation}
    q_{\mathrm{tot}} = f_1\left( T_{\mathrm{in}} - T_{\mathrm{out}}\right) - f_2(T_{\mathrm{out}})\Phi_{\mathrm{rad}} + q_{0},
\end{equation}
where $f_1$ and $f_2$ are unknown functions represented with chosen basis functions. We choose physically inspired ``flipped ReLU'' basis functions for $f_1$ that behave linearly below a given value and are zero otherwise. These functions are piecewise linear and account for the fact that the effect $f_1$ goes to zero above a certain value when the heating demand disappears. For $f_2$, we can assume that the effect is rather constant when there is heating demand and zero otherwise, which leads us to choose ``flipped step functions'' as the basis for $f_2$. These functions behave linearly between two grid points and are constant elsewhere. See Fig.~\ref{fig:gammy_basis} for an illustration of the chosen basis functions.

To account for cases with very little training data for the reference or intelligent-control periods, priors for the basis functions coefficient variance are given. This priors effectively regularize the fitted functions to not match for absurdly high total power which is assumed invalid data. In addition, the intelligent-control model $m_\mathrm{ic}$ uses a Gaussian prior for the value of the function $g_2$ at very cold outdoor temperature to match the value to the estimate of the reference model. This prior is useful for extrapolating the control effect outside the outside temperature range of the intelligent-control training data.

\begin{figure}[!ht]
    \centering
    \includegraphics[width=1.0\linewidth]{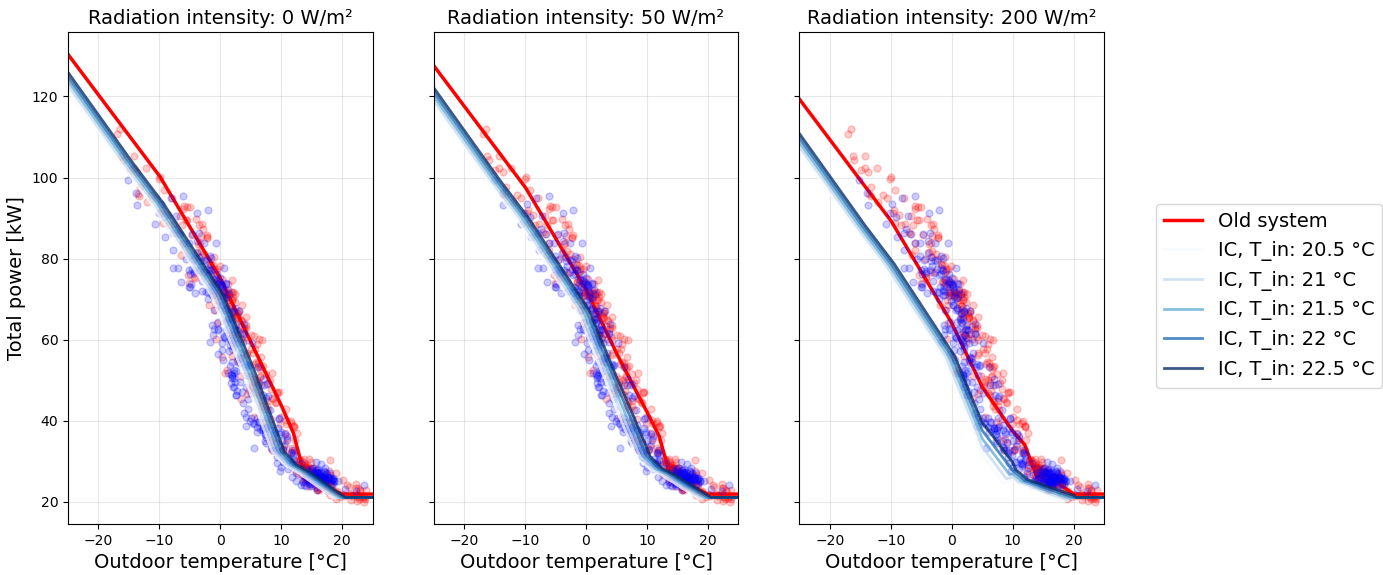}
    \caption{Data and GAM model predictions with varying indoor temperature and solar radiation levels.}
    \label{fig:gammy_fits}
\end{figure}

With these basis functions, we can calibrate Bayesian GAMs using data before and after the intelligent control system activation. An illustration of data and model predictions with various inputs is shown in Fig.~\ref{fig:gammy_fits}.

\section{HDD-based weather normalization details}
\label{sec:appendix_B}

As derived in Section \ref{sec:normalization}, the simplest HDD-based weather normalization is calculated as
\begin{equation}
    E_{\mathrm{tot}}^{\mathrm{ref}} = \frac{\mathrm{HDD}_{\mathrm{ref}}}{\mathrm{HDD}_{\mathrm{obs}}} \cdot (E_{\mathrm{tot}}^{\mathrm{obs}} - E_{\mathrm{dhw}}^{\mathrm{est}}) + E_{\mathrm{dhw}}^{\mathrm{est}},
\end{equation}
where $\mathrm{HDD}_{\mathrm{ref}} = \sum (T_{\mathrm{base}} - T_{\mathrm{out}}^{\mathrm{ref}})$ are the heating degree days of the reference year and $\mathrm{HDD}_{\mathrm{obs}} = \sum (T_{\mathrm{base}} - T_{\mathrm{out}}^{\mathrm{obs}})$ are the observed heating degree days at the building location.

To apply this in practice, one needs to decide three things: (i) how to set the reference indoor temperature $T_{\mathrm{base}}$, (ii) how to estimate the DHW consumption $E_{\mathrm{dhw}}^{\mathrm{est}}$, and (iii) how to deal with warm months when the heating demand is low, $\mathrm{HDD}_{\mathrm{obs}}$ is close to zero and the results are unstable. Here, we present one set of choices often employed in Finland due to the recommendation set by Motiva Oy, a Finnish government owned energy and sustainable development focused company.

In the Motiva method, the indoor reference temperature is set to $T_{\mathrm{base}}=17C$. In reality the apartments are warmer on average, but the remaining temperature difference is assumed to be the result of other energy sources, such as people, lighting, appliances, and solar radiation. The Motiva method offers multiple recommendations for domestic hot water consumption estimation. Preferred ways to determine the heating consumption for DHW require measuring the volume of consumed DHW, the volume of all consumed domestic water, or using an estimation based on the gross area of the building. If none of the previously mentioned options are viable, the heating consumption of DHW can be estimated to be the average total heating consumption of summer months, when the space
heating can be assumed to be off.

Finally, the dilemma with unstable results in the warm spring and autumn months is solved by discarding days with an average outdoor temperature of more than 10\,\textdegree C in the spring and more than 12\,\textdegree C in the autumn from the calculation. Even with these choices the results can be unstable, since intelligent control methods typically strongly reduce the heating level when the solar radiation is strong.

For more information on the Motiva normalization details, refer to \cite{rapo25} and \url{www.motiva.fi/kulutuksennormitus} (in Finnish).




\section{Examples of performance changes}
\label{sec:appendix_D}

In Section \ref{sec:changes}, we discussed various efficiency changes happening in the buildings that are not related to the control method. We gave an example of a ventilation system related change. Here, we give examples for some other change categories.

\begin{figure}[!ht]
    \centering
    \includegraphics[width=1.0\linewidth]{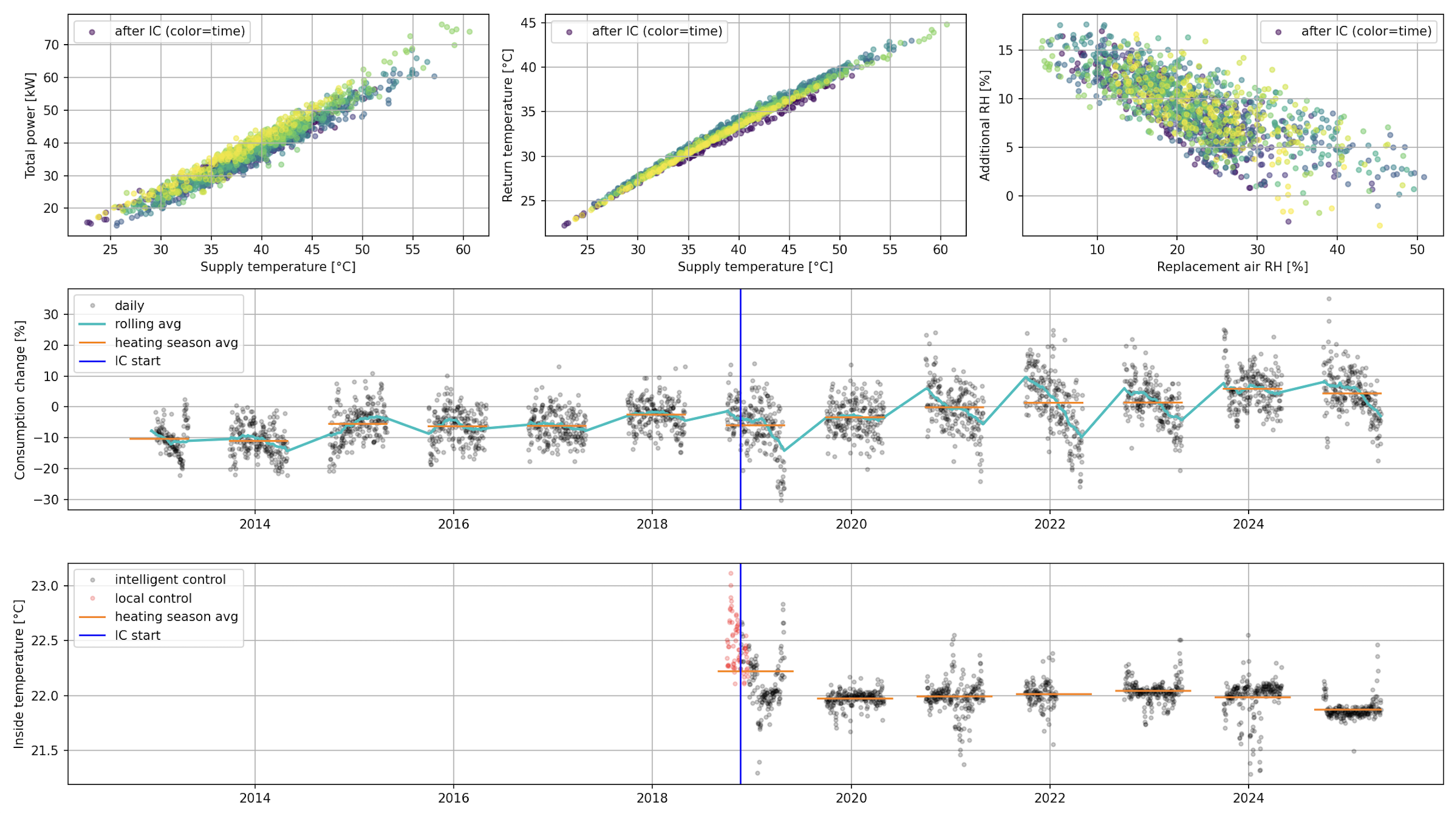}
    \caption{Example of a slowly worsening performance.}
    \label{fig:slow_changes}
\end{figure}

An example of a slow decrease in system performance is given in Fig.~\ref{fig:slow_changes}. No significant changes can be seen at the building dynamics or ventilation behavior, and indoor temperatures are kept at a constant level, but the consumption keeps increasing over time, probably due to aging of the building.

\begin{figure}[!ht]
    \centering
    \includegraphics[width=1.0\linewidth]{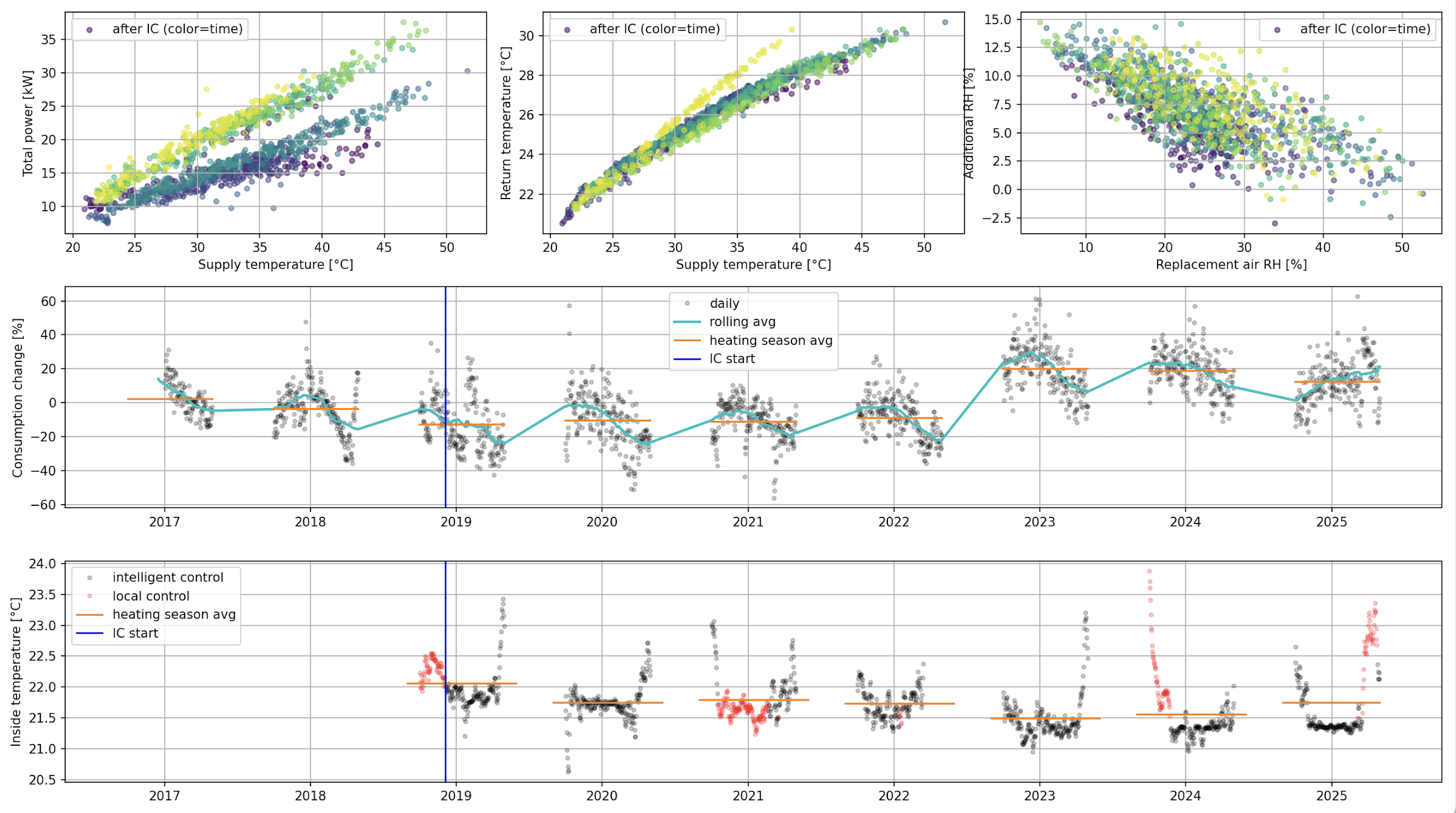}
    \caption{Example of a change happening in the heating system.}
    \label{fig:system_changes}
\end{figure}

Sometimes a change happens in the heating system that may affect the performance of the building. An example is given in Fig.\ref{fig:system_changes}. Here, the relationship between supply temperature and power consumption changes suddenly, and at the same time, the consumption increases significantly. The explanation is likely some operation done at the heating system level, but the exact reason remains unclear.

These examples highlight the fact that buildings undergo performance changes that are not related to the control method. This calls for performance tracking approaches that are not sensitive to these changes, which is the main contribution of this paper.

\bibliographystyle{unsrtnat}
\bibliography{references} 

\end{document}